\documentclass[aps,prm,superscriptaddress,reprint,showpacs,floatfix,pdftex]{revtex4-2}

\usepackage[whole]{bxcjkjatype}
\usepackage{graphicx}
\usepackage{comment}
\usepackage{tabularx}
\usepackage{dcolumn}
\usepackage{bm}
\usepackage{amsmath}
\usepackage{amssymb}
\usepackage{txfonts}
\usepackage{url}
\usepackage{xcolor}
\usepackage{multirow}
\usepackage{ulem}
\usepackage{subfigure}
\usepackage{physics}
\usepackage{braket}
\usepackage{mathrsfs}
\usepackage{empheq}
\newcommand{\textlabel}[1]{\label{#1}}
\newcommand{\mathlabel}[1]{\label{#1}}
\newcommand{\mycolor}{black}
\newcommand{\tadd}[1]{{\color{\mycolor}#1}}
\newcommand{\ttadd}[1]{{\color{\mycolor}#1}}

\newcommand{\fradd}[1]{{\color{\mycolor}#1}}
\newcommand{\jkadd}[1]{{\color{\mycolor}#1}}

\newcommand{\phname}{OPH23}

\begin{document}
\title{
  Locality Error Free Effective Core Potentials 
  for 3$d$ Transition Metal Elements 
  Developed for the Diffusion Monte Carlo Method
}
\thanks{
  Notice: This manuscript has been authored 
  by UT-Battelle, LLC, under
  contract DE-AC05-00OR22725 with the US Department of Energy (DOE).
  The US government retains and the publisher, by accepting
  the article for publication, acknowledges that the US
  government retains a nonexclusive, paid-up, irrevocable, worldwide
  license to publish or reproduce the published form 
  of this manuscript,
  or allow others to do so, for US government purposes.
  DOE will provide public access to these results
  of federally sponsored research in accordance 
  with the DOE Public Access Plan 
  (\url{http://energy.gov/downloads/doe-public-access-plan}).
}
\date{\today}
\author{Tom Ichibha {\textsuperscript{$a$}}}
\email[]{ichibha@icloud.com}
\affiliation{School of Information Science, JAIST, Asahidai 1-1, Nomi, Ishikawa 923-1292, Japan}

\affiliation{Materials Science and Technology Division, Oak Ridge National Laboratory, Oak Ridge, Tennessee 37831, USA}
\author{Yutaka Nikaido  {\textsuperscript{$a$}}}
\affiliation{School of Information Science, JAIST, Asahidai 1-1, Nomi, Ishikawa 923-1292, Japan}
\author{M. Chandler Bennett}
\affiliation{Materials Science and Technology Division, Oak Ridge National Laboratory, Oak Ridge, Tennessee 37831, USA}
\author{Jaron T. Krogel}
\affiliation{Materials Science and Technology Division, Oak Ridge National Laboratory, Oak Ridge, Tennessee 37831, USA}
\author{Kenta Hongo}
\affiliation{Research Center for Advanced Computing Infrastructure, JAIST, Asahidai 1-1, Nomi, Ishikawa 923-1292, Japan}
\author{Ryo Maezono}
\affiliation{School of Information Science, JAIST, Asahidai 1-1, Nomi, Ishikawa 923-1292, Japan}
\author{Fernando A. Reboredo}
\email[]{reboredofa@ornl.gov}
\affiliation{Materials Science and Technology Division, Oak Ridge National Laboratory, Oak Ridge, Tennessee 37831, USA}
\begin{abstract}
  Pseudopotential locality errors have hampered the applications 
  of the diffusion Monte Carlo (DMC) method 
  in materials containing transition metals, in particular oxides. 
  We have developed locality error free effective core potentials, 
  pseudo-Hamiltonians, for transition metals ranging 
  from Cr to Zn. We have modified a procedure published 
  by some of us in [M.C. Bennett et al, JCTC \textbf{18} (2022)]. 
  We carefully optimized our pseudo-Hamiltonians and achieved transferability errors comparable 
  to the best semilocal pseudopotentials used with DMC but without incurring in locality errors. 
    Our pseudo-Hamiltonian set (named  {\phname}) bears the potential to significantly 
    improve the accuracy of many-body-first-principles calculations in 
    fundamental science research of complex materials involving transition
    metals.
\end{abstract}
\maketitle
\textsuperscript{$a$} These two authors made the equivalent contributions.
\section{Introduction}
While initially \textit{ab initio} simulations were primarily 
used to provide physical insights on properties that are 
difficult to be observed experimentally \cite{2021TI_FAR}, or  
to verify and understand experimental data\cite{2022KO_FAR},
nowadays, they are increasingly being used 
to find new stable crystal structures \cite{2013AJ_KAP,2021TI_FAR}. 
This research aims to predict, before experiments, 
novel materials that have targeted physical properties.  
The importance of \textit{ab initio} simulations is 
therefore increasing in materials science. 
But the demand to predict a new stable material is 
also increasing the accuracy required in the results; 
which in turns demands improving the underlying theory. 

\vspace{2mm}
The diffusion Monte Carlo (DMC) method is emerging 
as one the most practical many-body \textit{ab initio} methods, 
because of its moderate cost scaling per step ($\sim N^3$) 
and excellent parallelizability 
\cite{2001FOU,2022YL_PK}.
In addition, DMC has comparable accuracy and reliability to 
the coupled cluster single-double and perturbative triple 
[CCSD(T)] method \cite{2012MJG_DA,2014BGAB_GQH,2013MD_LM}. 
DMC is often implemented within the fixed-node 
(or fixed-phase) approximation (FNDMC) 
to avoid the fermion sign problem 
\cite{1971RCG_RGS,1976JBA,2001FOU}. 
In FNDMC the nodes (phase) of the ground state wavefunction 
$\Psi_0$ are fixed by the ones given 
by trial wave function $\Psi_T$ provided by a different 
\textit{ab initio} method, like Kohn-Sham 
density functional theory (DFT) \cite{1965WK_LJS}. 
The fixed-node (phase) approximation introduced a positive bias 
in the total energy known as ``nodal errors''.
Nodal errors can be systematically reduced  
using multideterminant expansions, 
\cite{2012MAM_GES, 2020AB_MC} orbital optimization methods, 
or back-flow approximations
\cite{2006PLR_RJN,2013RM_RJN,2010RM_RJN}.

\vspace{2mm}
Accurate effective core potentials (ECPs) are also needed 
for practical problems.
ECPs are constructed so as they accurately reproduce 
multiple electronic 
scattering properties of all-electron (AE) calculations, 
often obtained at the frozen core approximation level.
In DMC the core electrons and ionic cores are replaced by an ECP 
to mitigate the computational cost.  
Without ECPs, the calculation cost of FNDMC scales as $Z^{5.5-6.5}$, 
where $Z$ is the nuclear charge \cite{1986DMC,1987BLH_WAL}. 
With ECPs, the cost scaling gets reduced to be 
$Z_{\mathrm{ECP}}^{3.4}$, where $Z_{\mathrm{ECP}}$ is 
the effective core charge \cite{1988BLH_WAL}.

\vspace{2mm}
The ECP approximation, however, brings a bias depending on 
the quality of ECPs (ECP errors) for any ab initio method
\cite{2008JK_RJN}. 
The typical semilocal form of ECP shown in Eq. \eqref{eq.semi} 
makes another type of errors (locality errors) when evaluated in FNDMC. 
Locality errors stem from the projection operators to 
the angular momentum channels: 
\tadd{$ \sum_{m=-\ell}^{\ell} \left | \ell m \right \rangle \langle \ell m | $}. 
\tadd{
  Here, $\left|\ell m\right>$ is the eigenstate for angular momentum $(\ell, m)$. 
}
Some approximation is needed to estimate the projections, 
\tadd{$\sum_{m=-\ell}^{\ell}  \Psi_T\left|\ell m\right>\left<\ell m\right|\Psi_0$},
avoiding an additional sign problem \cite{1998CM_DMC}. 
The conventional locality approximation \cite{1998CM_DMC} 
and the alternative $T$-moves \cite{2006MC,2010MS_CF} approximation 
can introduce errors in the range of 0.2--0.3 eV per atom
for the 3$d$ transition metal elements 
\tadd{
  for cohesive energies and relevant properties 
  \cite{2020JTK_FAR,2015JY_EE,2015JAS_FAR}. 
  The size of this error is roughly 10 times larger 
  than the remaining ECP's errors. 
  The state-of-the-art ECPs for DMC reach 0.02--0.03 eV errors 
  for the binding curves of 3$d$ transition metal and 
  oxygen dimers in CCSD(T) calculations \cite{2018AA_LM,2017MCB_LM}. 
  Therefore, the locality errors currently constitute the main contribution 
  of ECP errors in DMC applications to 3$d$ transition metal composites. 
}

\vspace{2mm}
Moreover, \tadd{these locality} errors are sometimes larger 
for the 4$d$ transition metal elements \cite{2016RN_RQH}. 
The magnitude of locality errors is dependent on the quality of 
the trial wavefunction $\Psi_T$, including the Jastrow factor 
\cite{Krogel2017,Dzubak2017}. 
Therefore, any comparison of results of different FNDMC calculations 
(required to predict magnetic properties, defect formation energies, 
band gaps etc.) becomes more dependent on the trial wave functions. 
Often trial wave functions have different locality errors 
since they are optimized\cite{2007UMR} independently.  
The uncertainty of the cancellation of locality errors
increases with the size of the electronic system considered. 
Accordingly, similar to the case of nodal errors, 
avoiding locality errors is a priority 
to improve the overall accuracy of the FNDMC approach. 

\vspace{2mm}
Another alternative, the auxiliary field quantum Monte Carlo 
(AFQMC) approach, \cite{2003SZ_HK,2006WAA_HK,2007MS_EJW,2013EP_US}
can avoid using pseudopotentials altogether 
with a frozen core approximation. 
Moreover, pseudopotentials can be incorporated in AFQMC 
without incurring in locality errors. 
AFQMC calculations are often one or two orders of magnitude 
more expensive than FNDMC. 
Therefore, alternative ways to accurately represent atomic cores, 
that do not include locality errors or increase significantly 
the computational cost, must be developed in FNDMC to compete 
in accuracy with AFQMC. 

\vspace{2mm}
A possible approach to avoid the locality errors in FNDMC 
is the pseudo-Hamiltonian (PH) framework \cite{1989GBB_MGBC}. 
Although the PH framework was originally proposed in 1989 \cite{1989GBB_MGBC}, 
\ttadd{
  this framework has been investigated only in a few cases
  \cite{1990WMCF_MS,1991XPL_RMM,1993AB_GBB}. 
  This is due to the numerical difficulty 
  to satisfy the constraint that 
  the eigenenergies of nodeless orbitals 
  increasing monotonically with the angular 
  momentum $\ell$ \cite{1990WMCF_MS}.  
  It is even harder to correctly describe 
  the scattering properties of multiple 
  angular-momentum channels together 
  along with the constraint \cite{1990WMCF_MS}.
  This constraint makes it difficult to construct
  an accurate PH, especially for some first row elements
  and Ar-core 3$d$ TMs\cite{2022MCB_JTK}.
}

\vspace{2mm}
\tadd{However,} the PH framework should be now reconsidered because recent progresses 
such as the multideterminant expansion \cite{2012MAM_GES,2020AB_MC} 
and Pfaffian pairing wave function 
\cite{2006MB_KES,2008MB_KES,2020CG_SS},
and increased computational resources have made  possible 
to reduce nodal and statistical errors in FNDMC so much that 
locality errors may become the leading source of error
in delicate energy sensitive comparisons involving transition metals. 
Therefore, some of us have developed a theoretical framework 
and procedures to construct PHs\cite{2020JTK_FAR,2022CB_JK} 
and demonstrated their effectiveness for the Co element 
\cite{2022CB_JK}. 

\vspace{2mm}
In this paper, we report newly constructed PH's for the 3$d$ 
transition metal elements from Cr to Zn with a slight modification 
of a previous procedure \cite{2022CB_JK}. 
We also compare the transferability of our PHs with 
highly accurate semilocal pseudopotentials used with FNDMC 
\cite{2007MB_MD,2008MB_MD,1984WJS_MK,1992WJS_PGJ,2018AA_LM,2017MCB_LM}.

\vspace{2mm}
The rest of the paper is organized as follows. 
In section \ref{sec.theory}, we explain the theory 
and the procedures used to generate multiple PHs 
\tadd{using \fradd{a} combination of Hartree-Fock (HF) and CCSD(T) methods} 
for each transition metal (TM) element.
\tadd{In Section \ref{sec.details}, 
  we explain the details of calculations conducted 
  for the optimization and verification of the PHs.}
In Subsection \ref{ssec.screening}, 
we explain how we selected the optimal PH\tadd{s} (named {\phname})
\tadd{based on the atomic properties and TM-O binding curves}
from the multiple candidate PHs we constructed for each TM element.
In Subsection \ref{ssec.comparison}, 
we confirmed \tadd{the reproducibility of AE}
binding curves of TM–O and TM–F molecules 
\tadd{
  by the {\phname} set, which is compared with 
} 
highly accurate semilocal pseudopotentials used with FNDMC,
BFD \cite{2007MB_MD,2008MB_MD}, SBKJC \cite{1984WJS_MK,1992WJS_PGJ}, 
and ccECP \cite{2018AA_LM,2017MCB_LM}. 
Finally, we summarize our study in Section \ref{sec.conclusion}.

\section{Theory}\textlabel{sec.theory}
\tadd{
The general form of pseudo-Hamiltonian \cite{1989GBB_MGBC} 
in spherical symmetry is given by: 
\begin{equation}
  \label{eq.psh} 
  \hat h^\mathrm{PH}(r) = \frac{1}{2}\hat{p}\left[1+a\left(r\right)\right]\hat{p} + {v}_\mathrm{local}^\mathrm{PH}(r) + v_{L^2}(r)\hat L^2.
\end{equation}
Here, $\hat p$ is the momentum operator.
The eigenenergies of nodeless orbitals necessarily 
monotonically increase with the angular momentum $\ell$ 
\cite{1989GBB_MGBC}. 
We choose to fix the radial mass to be constant, $a(r)=1$, 
as \fradd{in} the previous work by some of the authors \cite{2022MCB_JTK}. 
Then, the \fradd{first} term \fradd{on the right hand side of Eq. ({\ref{eq.psh}}) }comes down to the \fradd{usual} kinetic energy operator: 
\begin{equation}
\frac{\hat{p}\left[1+a\left(r\right)\right]\hat{p}}{2} \to \frac{\hat p ^2}{2}.
\end{equation}
This constraint is not required in theory 
but needed to \fradd{be able to} use general \textit{ab initio packages} with PHs 
because the variable radial mass is not currently implemented
in most codes used to obtain the trial functions for FNDMC. 
However, a sufficient transferability was achieved 
with this choice in the case of cobalt \cite{2022MCB_JTK}. 
}

\vspace{2mm}
Our \tadd{PH} is defined as
\tadd{
\begin{eqnarray}
  \hat v^\mathrm{PH}(r) &=& {v}_\mathrm{local}^\mathrm{PH}(r) + v_{L^2}(r)\hat L^2 
  \nonumber \\
  &=& {v}_\mathrm{local}^\mathrm{PH}(r)+ v_{L^2}(r) 
  \sum_{\ell=0}^{\infty} \sum_{m=-\ell}^{\ell} \ell (\ell + 1) {\left|\ell m\right>\left<\ell m\right|}. 
  \mathlabel{eq.ph}
\end{eqnarray}
  \fradd{The} $\left|\ell m\right>$ \fradd{are} the scalar spherical harmonics 
  as defined in Appendix A. 
}
This PH with the constant radial mass looks similar to the semilocal pseudopotentials 
\tadd{
\begin{equation}
  \hat v^\mathrm{SL}(r)=v_\mathrm{local}^\mathrm{SL}(r)
  + \sum_{\ell=0}^{M-1} \sum_{m=-\ell}^{\ell} v_\ell(r) { \left|\ell m\right>\left<\ell m\right|}. 
  \mathlabel{eq.semi}
\end{equation}
}
Here, $M-1$ is the maximum angular momentum channel 
of the core orbitals; $M$=2 for the neon core pseudopotentials 
for 3$d$ transition metal elements. 

\subsection{Initial values for the \tadd{PHs}}
Our strategy for PHs construction is to construct the parameter functions of the \tadd{PH}, 
$\hat{v}_\mathrm{local}^\mathrm{PH}(r)$ and $v_{L^2}(r)$, 
using the parameter functions of a successful semilocal pseudopotential as starting point.
For 3$d$ transition metal elements, we construct the starting $\hat{v}_\mathrm{local}^\mathrm{PH}(r)$ 
and $v_{L^2}(r)$ so as they reproduce the pseudopotential up to the $d$ channel:
\tadd{
\begin{align}
\label{eq.pshdef} 
\left<\ell m\left|\hat v\mathrm{^{PH}}\right|\ell m\right> 
&=\left<\ell m\left|\hat v\mathrm{^{SL}}\right|\ell m\right>,\\
(\ell=0,1,2;\;\;m&=-\ell, -\ell+1, \cdots, +\ell). \nonumber
\end{align}
}
\tadd{
Both sides \fradd{in Eq. ({\ref{eq.pshdef})}} are invariant with respect to $m$ as is clear from Eqs. \eqref{eq.ph} and \eqref{eq.semi}. 
Thus, there are actually three unique equations: 
}
\begin{eqnarray}
v_\mathrm{local}^\mathrm{PH}(r) &=& v_\mathrm{local}^\mathrm{SL}(r) + v_0(r), 
\mathlabel{eq.cond1} \\
v_\mathrm{local}^\mathrm{PH}(r) + 2v_{L^2}(r) &=& v_\mathrm{local}^\mathrm{SL}(r) + v_1(r),
  \mathlabel{eq.cond2} \\
  v_\mathrm{local}^\mathrm{PH}(r) + 6v_{L^2}(r) &=& v_\mathrm{local}^\mathrm{SL}(r).
  \mathlabel{eq.cond3} 
\end{eqnarray}
Equations \eqref{eq.cond1}--\eqref{eq.cond3} can 
be solved imposing  a condition of linear dependence:
\begin{equation}
  2 v_0(r) - 3 v_1(r) = 0.
  \mathlabel{eq.linDep1}
\end{equation}
To facilitate the optimization, we choose to modify $v_1(r)$ as the function of $v_0(r)$ 
\begin{equation}
  v_1[v_0(r)] = (2/3)v_0(r), \mathlabel{eq.linDep2}
\end{equation}
because the semilocal 3$p$ 
orbitals are \tadd{ 
\fradd{for transition metals} much} less important 
than the valence 3$d$ and 4$s$ orbitals to describe 
chemical reactions \cite{2018AA_LM,1987MD_HP}.
As a result, the starting parameter functions of PH 
are given from Eqs. \eqref{eq.cond1}--\eqref{eq.cond3} 
and Eq. \eqref{eq.linDep2} as 
\begin{equation}
{v}_{\mathrm{local}}^{\mathrm{PH}}
=v_{\mathrm{local}}^{\mathrm{SL}}+v_0, \;
v_{L^2}=-(1/6)v_0.
\end{equation}

\vspace{2mm}
In addition, the parameter function $v_{L^2}$ has to satisfy the following bounding condition
\cite{2022MCB_JTK, 2020JTK_FAR}: 
\begin{eqnarray}
  1+2r^2v_{L^2}(r) > 0. \mathlabel{eq.bound}
\end{eqnarray}
The left-hand side in Eq. (\ref{eq.bound}) corresponds to the value of an effective mass 
multiplying the electron angular momentum. 
A negative effective mass would lead to an electronic wavefunction to localize into a delta function lowering its energy. 
Accordingly, Eq. \eqref{eq.bound} must be satisfied to avoid unphysical results. 
Therefore, we replace $v_{L^2}(r)$ with a smooth and shifted 
ramp function $v'_{L^2}(r)$ that satisfies the bounding condition: 
\begin{eqnarray}
  2r^2 v_{L^2}(r) \to 2r^2 v'_{L^2}(r) \equiv \mathcal{R_S}\left(2r^2v_{L^2}(r), \delta b, \delta b_s  \right).
  \mathlabel{eq.ramp}
\end{eqnarray}
The ramp function $\mathcal{R_S}(x)$ is shown in Figure \ref{fig.rampFunc}.
When $x=2r^2v_{L^2}(r)$ approaches the lower bound $-1$ from the positive side,
$\mathcal{R_S}(x)$ smoothly converges to $-1+\delta b_s$; 
$2r^2 v'_{L^2}(r)$ always satisfy the condition \eqref{eq.bound}. 
When $x=2r^2v_{L^2}(r)$ becomes much larger than the lower limit $-1$,
$\mathcal{R_S}(x)$ smoothly converges to $x=2r^2v_{L^2}(r)$.
$\delta b$ decides the width of connection region
between the constant and proportional regions in $\mathcal{R_S}$.
The mathematical details of $\mathcal{R_S}$ can be found
in previous works \cite{2022MCB_JTK, 2020JTK_FAR}.
  
\vspace{2mm}
The difference between $2r^2 v_{L^2}(r)$ and $2r^2 v'_{L^2}(r)$:
\begin{equation}
  \Delta(r) \equiv 2v'_{L^2}(r) - 2v_{L^2}(r) \mathlabel{eq.delta}
\end{equation}
was found to be well fitted by Gaussian functions
in the case of cobalt \cite{2022MCB_JTK}: 
$\Delta(r)$ with $\mathcal{R_S}\left(2r^2v_{L^2}(r), 
\delta b=1.6\,\mathrm{Ha}, \delta b_s=5 \times 10^{-4}\,\mathrm{Ha} \right)$
was expanded by three Gaussian primitives with exponent values being 
12.5, 18.75, and 28.128 Bohr$^{-2}$\cite{2022MCB_JTK}.
In this work, we also expand $\Delta(r)$ in the same manner for the 3$d$ transition elements.
Eventually, we replace $2r^2 v_{L^2}(r)$ as 
\begin{equation}
2r^2 v_{L^2}(r) \to 2r^2v''_{L^2}(r) \equiv 2r^2 v_{L^2}(r) + r^2 \Delta_{\mathrm{expanded}}(r).
\end{equation}
From these initial values, the coefficients and coordinates of the Gaussian functions are optimized along with ${\hat v}_{\mathrm{local}}^{\mathrm{PH}}$ and $v_{L^2}$. 

\begin{figure}[htbp]
  \centering
  \includegraphics[width=\hsize]{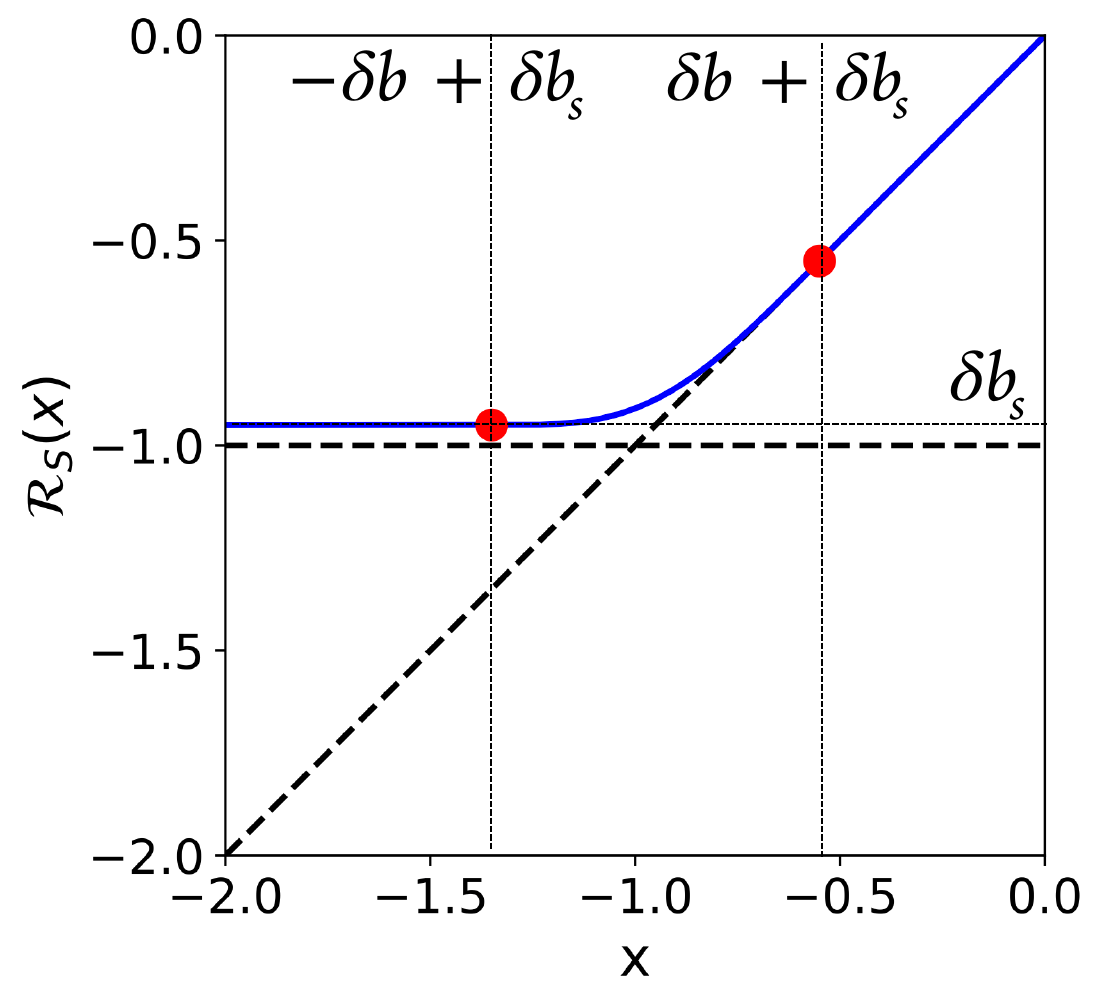}
  \caption{\textlabel{fig.rampFunc}
    Plot of the smooth and shifted ramp function. 
    This figure is a reprint of Fig. 8 of our previous 
    work \cite{2020JTK_FAR} with small modification. 
  }
\end{figure}

\subsection{Pseudo-Hamiltonian \tadd{construction}}
\label{ssec.phopt}

\tadd{
  We \fradd{choose} 
  the BFD semilocal pseudopotentials to define 
  the \fradd{initial} parameter functions of our PHs 
  via Eqs. \eqref{eq.cond1}--\eqref{eq.cond3}
  because of the\fradd{ir} simple parameterization. 
  The core size of our PHs is the neon core 
  as well as BFD, SBKJC, and ccECP. 
  Ne-core pseudopotentials have been found 
  to have better transferability than Ar-core 
  in general with FNDMC method \cite{1987MD_HP,1996DMC_LM,1994LM}. 
}

\vspace{2mm}
We optimize the parameters in $\hat v\mathrm{^{PH}}$ via three steps. 
In the first step, we conduct optimization to 
\tadd{
  reproduce all the valence orbital norms and eigenenergies as well as 
  neutral excitation and ionization energy spectrum in 
  4$s$ and 3$d$ levels \fradd{obtained with}  the HF method 
  with SBKJC pseudopotential.  
}
In the second step, we select the \tadd{PH} that best reproduces 
the TM–O binding curve \tadd{among the multiple PHs for each element}.
In the third step, we further optimize the selected \tadd{PHs} 
so as to reproduce \tadd{all} the valence orbital norms and eigenenergies 
at the HF level and energy spectrum at the CCSD(T) level with ccECP pseudopotential. 
\tadd{
  The details of objective functions and procedures are 
  described in the following subsections.
}


\subsubsection{Hartree-Fock optimization} 
\label{ssec.hfopt}

\begin{table}[htbp]
  \centering
  \caption{
    \label{tab.state}
    \tadd{
      Reference electronic states for optimizing 
      the PH of each element. 
    }
  }
  \begin{tabular}{ccccccc}
    \hline
    Cr & Mn & Fe & Co & Ni & Cu & Zn \\
    \hline
    4$s^{1}$3$d^{5}$ &  4$s^{2}$3$d^{5}$ & 4$s^{2}$3$d^{6}$ & 4$s^{2}$3$d^{7}$
    & 4$s^{1}$3$d^{9}$ & 4$s^{1}$3$d^{10}$ & 4$s^{2}$3$d^{10}$ \\
    \hline    
  \end{tabular}
\end{table}

\fradd{The reference electronic states used to optimize the PHs in HF are listed in Table \ref{tab.state}.  
  In general, we select the atomic ground state as the reference state. 
  However, for the nickel case, because we could not obtain a solution 
  that satisfies the bounding condition in Eq. \eqref{eq.bound} 
  with the ground state 4$s^{2}$3$d^{8}$, 
  we select the nearly degenerated state 4$s^{1}$3$d^{9}$ 
  as the reference state.} 

\vspace{2mm}
\fradd{For each element,} we optimize the pseudo-Hamlitonian to minimize 
the following cost function: 
\begin{equation}
  \mathcal{O}^{\mathrm{HF}} = \omega_1 \mathcal{O}_{\mathrm{Eigenvalue}} + \omega_2 \mathcal{O}_{\mathrm{Norm}}
  + \omega_3 \mathcal{O}_{\mathrm{Bound}} + \omega_4 \mathcal{O}_{\mathrm{Excitation}}^{\mathrm{HF,SBKJC}}, 
  \mathlabel{eq.cost}
\end{equation}
where,
\begin{equation}
  \label{eq.eigenval}
  \mathcal{O}_{\mathrm{Eigenvalue}} = \sum_i \left[\left(\varepsilon_i^{\mathrm{{PH}}} - \varepsilon_i^{\mathrm{SBKJC}}\right) \middle/ \varepsilon_i^{\mathrm{SBKJC}} \right]^2,
  \end{equation}
  \begin{equation}
  \label{eq.norm} 
  \mathcal{O}_{\mathrm{Norm}} = \sum_i \left[\braket{\phi_i^{\mathrm{PH}}|\phi_i^{\mathrm{PH}}}_{r>0.8a_0} - \braket{\phi_i^{\mathrm{SBKJC}}|\phi_i^{\mathrm{SBKJC}}}_{r>0.8a_0} \right]^2,
  \end{equation}
  \begin{equation}
  \mathcal{O}_{\mathrm{Bound}} = \int dr \left[\mathcal{R_S}\left(2r^2v''_{L^2}\left(r\right), \delta b'
  , \delta b'_s \right)-2r^2 v''_{L^2}\left(r\right)  \right],
  \end{equation}
  \begin{equation}
  \mathcal{O}_{\mathrm{Excitation}}^{\mathrm{HF,SBKJC}} =  \sum_{j\ni\left(3d,4s\right)} \left[\frac{ \Delta E_j^{\mathrm{HF,PH}} - \Delta E_j^{\mathrm{HF,SBKJC}}}{\Delta E_j^{\mathrm{HF,SBKJC}}} \right]^2.
  \mathlabel{eq.costgap}
\end{equation}
\tadd{
  $\sum_i$ indicates a summation over all the valence orbitals.
  $\sum_{j\ni\left(3d,4s\right)}$ indicates a summation over excited and ionized electronic 
  states in 3$d$ and 4$s$ levels. 
  Further explanation is found in the last paragraph of this subsubsection. 
}
$\braket{}_{r>0.8a_0}$ indicates 
an integral outside of the 0.8 $a_0$ ($a_0$: Bohr radius) 
radius sphere around the core center. 
$\mathcal{O}_{\mathrm{Eigenvalue}}$ is lower 
when the \tadd{orbital eigenenergies} are reproduced better.
$\mathcal{O}_{\mathrm{Norm}}$ is lower when the integration
of \tadd{each orbital} norm outside of the 0.8 $a_0$ cutoff 
is reproduced better.

\vspace{2mm}
\jkadd{The starting PH used for subsequent optimization initially alters the shape of the p-channel found in the reference ECP (see Eq. (\ref{eq.linDep2})).  During PH optimization, the quality of the p-channel is restored--in balance with the others--since $3p$ orbitals are included in the cost function in Eqs. (\ref{eq.eigenval}) and (\ref{eq.norm}).
In 3d transition metals},  \fradd{the 3p orbitals are  highly localized, deep and inert.  Enforcing the norm and eigenvalues for the $3p$ \jkadd{subshell has been shown to be sufficient} to obtain accurate pseudopotentials \jkadd{for 3d transition metals}.} \cite{2019GW_LM}.

\vspace{2mm}
$\mathcal{O}_{\mathrm{Bound}}$ is larger when the bounding 
condition \eqref{eq.bound} approaches being violated, 
and is otherwise almost zero. 
We employ $\delta b'=0.25\,\mathrm{Ha}$ and 
$\delta'_s = 5 \times 10^{-3}\,\mathrm{Ha}$
as our previous work \cite{2022MCB_JTK}.

\vspace{2mm}
\tadd{
$\mathcal{O}_{\mathrm{Excitation}}^{\mathrm{HF,SBKJC}}$ is 
  lower when the energy spectrum of neutral excitation and 
  ionization in the 3$d$ and 4$s$ levels are reproduced better.
} 

\vspace{2mm}
\tadd{ 
The electron transitions in 3$d$ and 4$s$ orbitals 
  and \fradd{a} variety of oxidation states are essential 
  for the chemistry of 3$d$ transition metal\fradd{s} \cite{2018AA_LM,1987MD_HP}; thus they are included in Eq.(\ref{eq.costgap}).
  \fradd{
  However, following a proven procedure in the literature for TM pseudo potentials\cite{2019GW_LM}, we} 
  \fradd{do not consider}
  the 3$p$ \fradd{to} 4$p$ levels excitation in the energy spectrum \fradd{in Eq. (\ref{eq.costgap})}. 
}
  
\vspace{2mm}
\jkadd{Additionally, it is known that}
  \fradd{
  TM-O dimers strongly hybridize  s,p,d orbitals.
  Therefore, since excitations to $4p$ are not weighted in Eq. (\ref{eq.costgap}), we performed additional test
calculations of TM-O dimer binding curves to assess the transferability of our PHs (See \ref{ssc.binding}). }  


\vspace{2mm}
There are still some degrees of freedom to choose: 
which excited and ionized states are considered in the evaluation 
of Eq. \eqref{eq.costgap}. 
We take into account every possible combination of 
3$d$ and 4$s$ electrons from the charge neutral to $+X$ charged states,
where the electron configuration is decided by the Hund's rules.
We sometimes ignore a small number of excited states 
which cannot be calculated 
(when  either the HF or CCSD(T) calculations do not converge).  
We optimize a \tadd{PH} for every $X$=2--6.
We denote them in the following sections as HFOPT($X$), respectively. 

\subsubsection{Binding Curve Selection \label{ssc.binding}}
For each TM, we select a HFOPT($X$) that best reproduces 
the molecular TM--O binding curve given by CCSD(T) 
in order to further optimize it at the third step.

\subsubsection{Many-body correction}
\label{ssec.m-bc}
We make minor and major modifications to the cost function 
in Eq. \eqref{eq.cost} and further optimize the selected 
\tadd{PH} to minimize the modified cost function.
The minor modification is using the ccECP pseudopotentials 
\cite{2018AA_LM,2017MCB_LM} instead of the SBKJC pseudopotentials 
to provide the reference HF results; 
we found that ccECP pseudopotentials tend to reproduce slightly 
better the AE CCSD(T) TM–O binding curves. 
The major modification we made is to replace 
$\mathcal{O}_{\mathrm{Excitation}}^{\mathrm{HF,SBKJC}}$ 
in Eq. \eqref{eq.costgap} with 
\begin{equation}
  \mathcal{O}_{\mathrm{Excitation}}^{\mathrm{CCSD(T),ccECP}}
  = \sum_{j\ni\left(3d,4s\right)} \left(
    \frac{\Delta E_j^{\mathrm{HF,CCOPT}}-\Delta E_j^{\mathrm{CC}}}
         {\Delta E_j^{\mathrm{CC}}}
         \right)^2, 
         \mathlabel{eq.ccopt} 
\end{equation}
         \begin{equation}
  \Delta E_j^{\mathrm{CC}} \equiv \Delta E_j^{\mathrm{HF,HFOPT}} - \Delta E_j^{\mathrm{CCSD(T),HFOPT}} 
  + \Delta E_j^{\mathrm{CCSD(T),ccECP}}
  \mathlabel{eq.cc}
\end{equation}
\cite{2022CB_JK}. 
Here, HFOPT is the \tadd{PH} that we select at the second step. 
CCOPT is the \tadd{PH} we optimize at this third step. 
The minimization of cost function in Eq. \eqref{eq.ccopt} 
corresponds to the optimization so as to reproduce the electron 
excitation energies at the CCSD(T) level 
\tadd{(see Appendix B)}. 
The advantage of Eq. \eqref{eq.ccopt} is not including 
$\Delta E_i^{\mathrm{CCSD(T),CCOPT}}$,
which requires expensive CCSD(T) calculations 
for hundreds of parameter updates. 

\vspace{2mm}
We compare the best HFOPT and CCOPT in terms of reproducibility of 
AE TM–O binding curve and AE atomic ionization energies. 
We name the better \tadd{PH} {\phname}, 
the \textbf{o}ptimal \textbf{p}seudo\textbf{H}amiltonian produced in 20\textbf{23}.
We compare the {\phname}'s reproducibility of AE TM–O and AE TM–F binding curves with 
highly accurate semilocal pseudopotentials used with FNDMC,
BFD\cite{2007MB_MD,2008MB_MD}, SBKJC\cite{1984WJS_MK,1992WJS_PGJ}, and ccECP\cite{2018AA_LM,2017MCB_LM}
in Section \ref{sec.results}.

\section{Calculation details}
\textlabel{sec.details}
We used libraries in the Nexus workflow management software\cite{2016KRO} 
to optimize the \tadd{PH}s and convert ECP file format. 
We employed the Nelder--Mead algorithm\cite{1965JAN_RM,1962WS_FRH} 
implemented in SciPy\cite{2020PV_S1C} for parameter fitting of \tadd{PH}s. 
During the optimization process, we used a numerical 
atomic Hartree--Fock code 
\cite{1997SK_CWCa,1997SK_CWCb} 
to evaluate the eigenengies, norms, and  excitation energies of atoms. 
We used the Molpro code\cite{2011HJW_MS,2020HJW_MS} to evaluate 
the $\Delta E_j^\mathrm{CC}$ term of Eq. \eqref{eq.cc} and 
the dimer binding curves and atomic excitation energies with CCSD(T).
We evaluated the dimer binding curves with our \tadd{PH}s and 
semilocal pseudopotentials used with FNDMC: BFD\cite{2007MB_MD,2008MB_MD}, 
SBKJC\cite{1984WJS_MK,1992WJS_PGJ}, and ccECP\cite{2018AA_LM,2017MCB_LM}, 
for comparison, and by AE calculations as the reference.
Here, we truncated the $\ell > 7$ terms of \tadd{PH}s in Eq. \eqref{eq.ph} because Molpro cannot deal with $\hat L^2$ directly. 
We have confirmed that the binding curve sufficiently converges with this truncation in the case of Co--O. 
We used the 10th order Douglas--Kroll--Hess relativistic 
Hamiltonian\cite{1974MD_NMK,1986BAH} for the AE calculations.
We used the non-relativistic Hamiltonian for calculations with ECPs. 

\vspace{2mm}
\tadd{
We evaluated the ionization energies shown in Figures \ref{fig.screening.1} and \ref{fig.screening.2} with the complete basis set (CBS) limit calculated with the fully uncontracted aug-cc-pwCV$n$Z-DK basis sets ($n$=T, Q, and 5) \cite{2006NBB_KAP}. The CBS limit was determined according to the methodology outlined in the ccECP paper for 3$d$ TM elements \cite{2018AA_LM}. Specifically, the Hartree-Fock (HF) energy was extrapolated using
\begin{equation}
E_{n}^{\mathrm{HF}}=E_{\mathrm{CBS}}^{\mathrm{HF}}+a\exp\left[-bn\right].
\end{equation}
Similarly, the correlation energy, defined as the difference between CCSD(T) and HF energies, was extrapolated using
\begin{equation}
E_{n}^{\mathrm{corr}}=E_{\mathrm{CBS}}^{\mathrm{corr}}+\frac{c}{\left(n+3/8\right)^3}+\frac{d}{\left(n+3/8\right)^5}.
\end{equation}
To calculate binding curves and $\Delta E_i^{\mathrm{CC}}$ in Eq. \eqref{eq.ccopt}, 
we utilized the fully uncontracted aug-cc-pwCVQZ-DK basis sets \cite{2006NBB_KAP}. 
For the {\phname} binding curve relative to the AE, 
the deviations from the CBS limit were found to be less than 7 meV,
in the case of Co--O dimer. 
Previous research \cite{2018AA_LM} has also suggested that the quadruple zeta level 
is adequate to calculate such 
binding curves. 
Deviations from the CBS limit in $\Delta E_i^{\mathrm{CC}}$ 
were below 1.8 \%, which is lower than the discrepancies observed between ccECP 
and AE excitation energies \cite{2022MCB_JTK}. 
Data for the error estimations are provided in the Supporting Information.
}

\section{Results and discussion}
\textlabel{sec.results}
\subsection{Screening of \tadd{PH}s}
\textlabel{ssec.screening}
In this Subsection, we discuss (1) how we selected 
the optimal HFOPT($X_{\mathrm{opt}}$) from HFOPT($X$) 
($X=$2--6) and (2) how we determined whether to employ 
HFOPT($X_{\mathrm{opt}}$) or CCOPT($X_{\mathrm{opt}}$) 
as the {\phname}. 
As a reminder for the discussion below, PH names post-fixed 
with ($X$) denote potentials that were constructed 
\tadd{
  to reproduce the energy spectrum with neutral excitation 
  and ionization up to +$X$ charge in 3$d$ and 4$s$ levels. 
}

\vspace{2mm}
The comparisons of the PHs are shown in Figure \ref{fig.screening.1} 
for each element from Cr to Co, and in Figure \ref{fig.screening.2} 
for each element from Ni to Zn.
The vertical dashed and dotted lines in the left column indicate
respectively the AE binding distance given by AE calculations 
and TM–O closest distance in the reference transition metal oxide 
listed in Table \ref{tab.distance}. 
Accurate binding observed in this range is suitable to qualify a 
particular PH for use in target solid-state applications.
\tadd{The AE binding distances were obtained by the cubic spline interpolation.}

\vspace{2mm}
\textit{Chromium:} 
Every HFOPT reproduces the AE binding curve given 
by CCSD(T) within the chemical accuracy. 
HFOPT(4), HFOPT(5), and HFOPT(6) similarly give 
the lowest deviations. 
We made the CCOPT for $X$=4. 
Although CCOPT(4) does not reproduce the AE binding curve 
within the chemical accuracy in the very short distances 
unlike HFOPT(4), CCOPT(4) much better reproduces 
the excitation energy (4s$^1$3d$^5$ $\to$ 4s$^2$3d$^4$) 
than HFOPT(4). We selected CCOPT(4) as the {\phname}.

\vspace{2mm}
\textit{Manganese:}
Every HFOPT similarly reproduces the AE binding curve almost 
within the chemical accuracy. 
HFOPT(5) gives the smallest deviations 
around the AE binding distance
so we made the CCOPT for $X$=5. 
CCOPT(5) gives larger deviations for the binding curve 
than HFOPT(5).
We selected HFOPT(5) as the {\phname}. 

\vspace{2mm}
\textit{Iron:}
The HFOPTs give different binding curves. 
HFOPT(5) gives the smallest deviations among the HFOPTs. 
However, the chemical accuracy is not satisfied 
for the shorter distances. 
We made the CCOPT for $X$=5. 
CCOPT(5) reproduces the AE binding curve 
within the chemical accuracy 
and better reproduces the AE \tadd{energy spectrum} 
compared with HFOPT(5). 
We selected CCOPT(5) as the {\phname}. 

\vspace{2mm}
\textit{Cobalt:}
The HFOPTs give similar binding curves and do not satisfy 
the chemical accuracy for the shorter distances. 
CCOPT(3) gives the smallest errors for the shorter distances. 
We made the CCOPT for $X$=3. 
CCOPT(3) reproduces the AE binding curve within the chemical accuracy 
and better reproduces the AE \tadd{energy spectrum} compared with HFOPT(3). 
We selected CCOPT(3) as the {\phname}. 

\vspace{2mm}
\textit{Nickel:}
Every HFOPT similarly reproduces the AE binding curve 
within the chemical accuracy. We made the CCOPT for $X=$5. 
CCOPT(5) does not reproduce the AE binding curve within the chemical accuracy 
for the shorter distances and give larger error for 
\tadd{the excitation from $4s^{1}3d^{9} \to 4s^{2}3d^{8}$}.
We selected HFOPT(5) as the {\phname}. 
\tadd{
  The remarkably large percentage error for the excitation 
  is due to the very low excitation energy  ($\sim$20 meV).
  Thus, 10 \% error actually corresponds to only 2 meV. 
}

\vspace{2mm}
\textit{Copper:}
HFOPT(2) (outside of the display range) and HFOPT(3) gave 
very large deviations for the binding curve. 
HFOPT(4), HFOPT(5), and HFOPT(6) reproduces the AE binding curve 
within chemical accuracy for the longer distances than 
the Cu–O closest distance in CuO solid. 
HFOPT(4) gives the lowest error within the HFOPTs so we made the CCOPT for $X=$4. 
However, CCOPT(4) does not reproduce the AE binding curve. 
We selected HFOPT(4) as the {\phname}. 

\vspace{2mm}
\textit{Zinc:}
The HFOPTs give very different binding curves depending on $X$. 
Among them, HFOPT(3) gives constant deviations from the AE binding curve 
(i.e., good reproducibility of curvature), so we choose the CCOPT for $X=$3.
CCOPT(3) shows slightly smaller deviations for the binding curve
and excellent reproducibility of the AE \tadd{energy spectrum}. 
We selected CCOPT(3) as the {\phname}. 

\vspace{2mm}
Note, with the exception of Mn, the other \tadd{PH}s are within chemical accuracy 
for the typical oxide's distances listed in Table \ref{tab.distance}. 
In contrast, the accuracy of the {\phname} set tends to decrease for shorter distances.

\vspace{2mm}
In summary, we selected CCOPT($X_\mathrm{opt}$) as the {\phname} 
for Cr, Fe, Co, and Zn and HFOPT($X_\mathrm{opt}$) for Mn, Ni, and Cu. 

\begin{table}[htbp]
  \centering
  \caption{
    \textlabel{tab.distance}
	The TM–O and TM-F closest distances (\AA) in transition metal oxides and fluorides.
  }
  \scalebox{1.0}[1.0]{
  \begin{tabular}{cccc}
    \hline
	TM–O name & TM-O distance & TM–F name & TM-F distance \\
    \hline
	  Cr$_2$O$_3$ & 2.012  \cite{2012AK_LD} & CrF$_2$ & 1.978  \cite{1963WG} \\
    MnO & 2.222  \cite{1991REP_EKG} & MnF$_2$ & 2.128  \cite{1978HM_FW} \\
	  FeO & 2.154  \cite{2002HF_SS} & FeF$_2$ & 2.118  \cite{1971BH_KA} \\
    CoO & 2.132  \cite{1979SS_YT} & CoF$_2$ & 2.058  \cite{1993CR_AR} \\
	  NiO & 2.072  \cite{1998HT} & NiF$_2$ & 2.011  \cite{1993CR_AR} \\
	  CuO & 1.936  \cite{1996NJC_GBS} & CuF$_2$ & 1.917  \cite{1974FP_GH} \\
	  ZnO & 1.974  \cite{1985KK_DG} & ZnF$_2$ & 2.041  \cite{2001ON_SA} \\
    \hline
  \end{tabular}
  }
\end{table}
\newcommand{\scale}{0.36}
\newcommand{\offset}{2.5cm}
\newcommand{\vertical}{-0.7cm}
\begin{figure*}[htbp]
  \centering  
  \begin{tabular}{rcc}
	& {\Large TM-O binding} & {\Large Ionization} \\
    \raisebox{\offset}{\LARGE Cr} &
    \includegraphics[width=\scale\hsize]{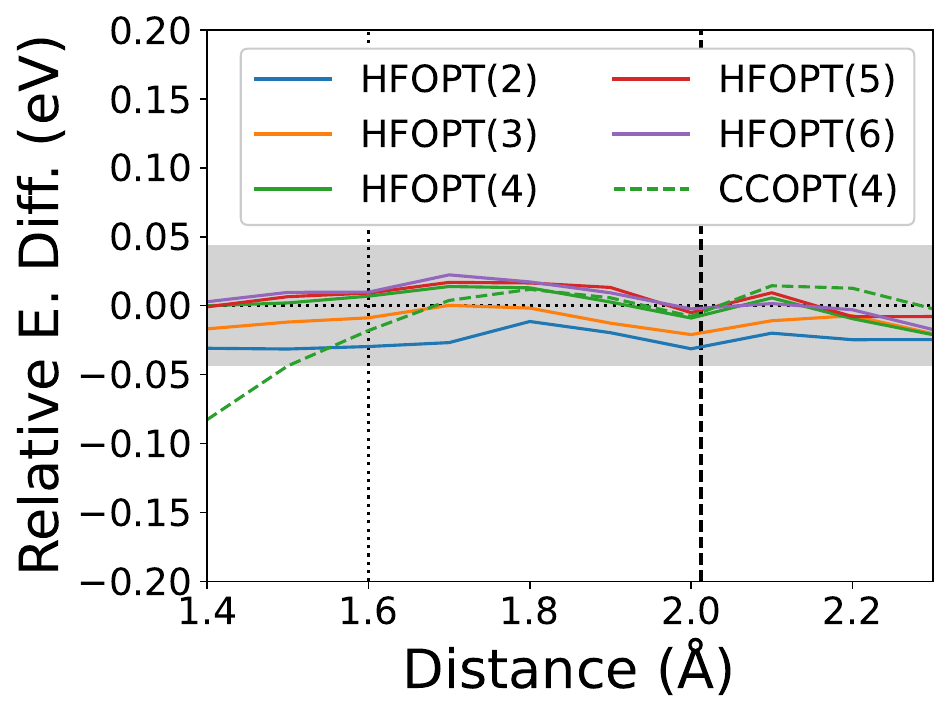} &	
    \includegraphics[width=\scale\hsize]{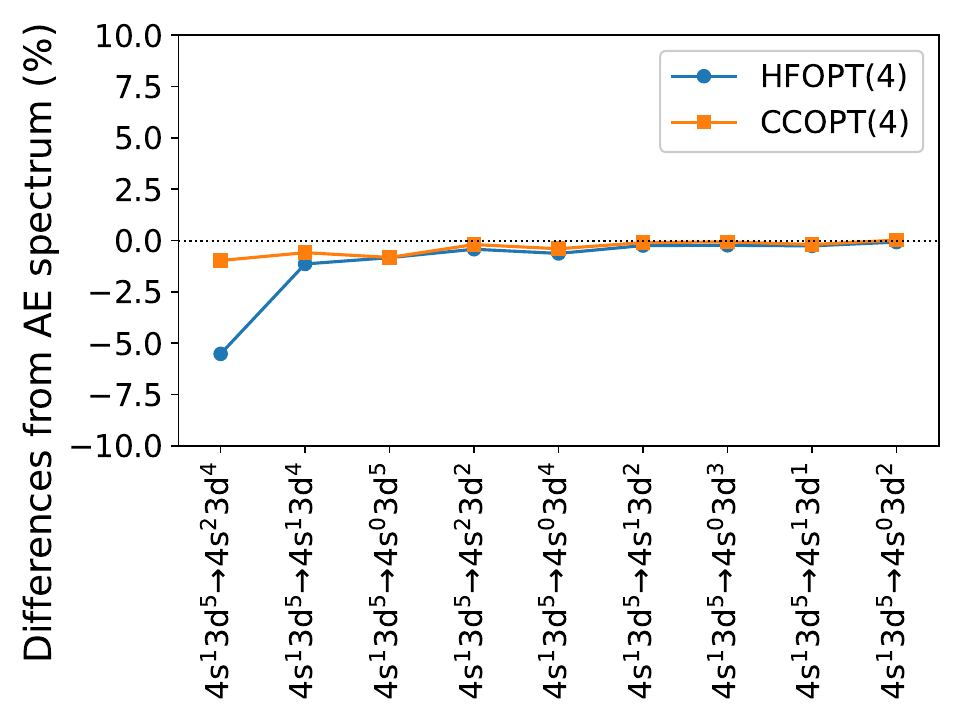} \\
    \raisebox{\offset}{\LARGE Mn} &
    \includegraphics[width=\scale\hsize]{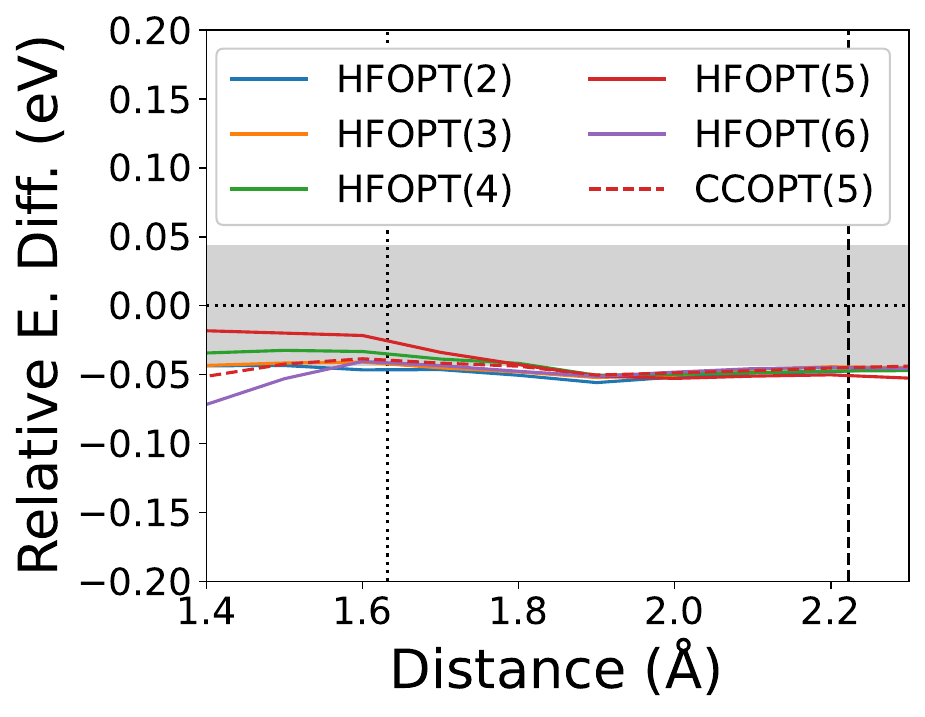} &
    \includegraphics[width=\scale\hsize]{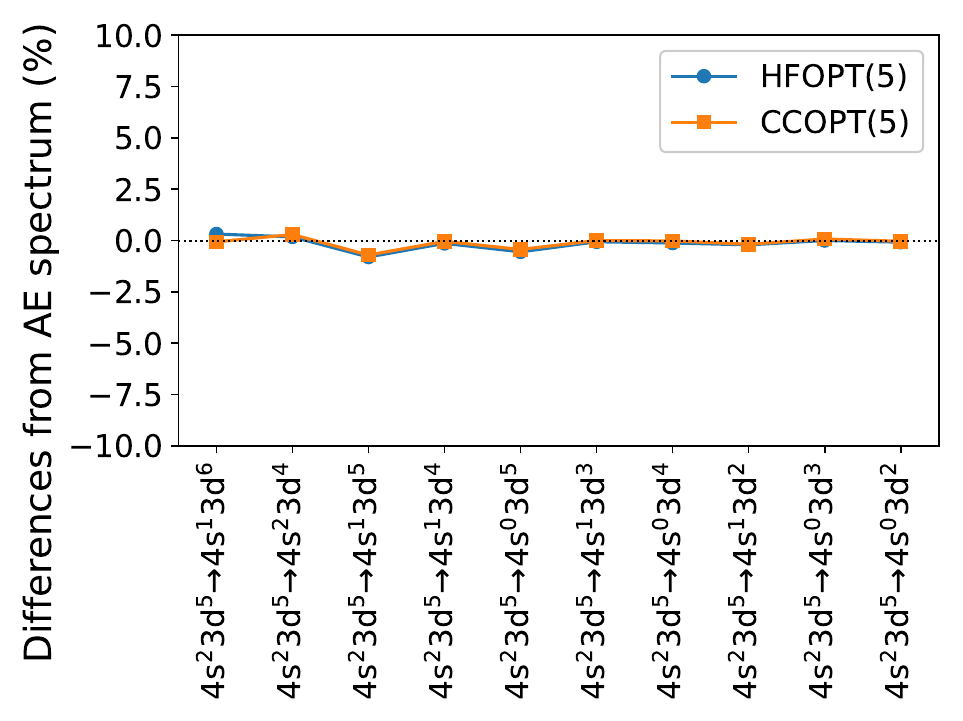} \\
    \raisebox{\offset}{\LARGE Fe} &
    \includegraphics[width=\scale\hsize]{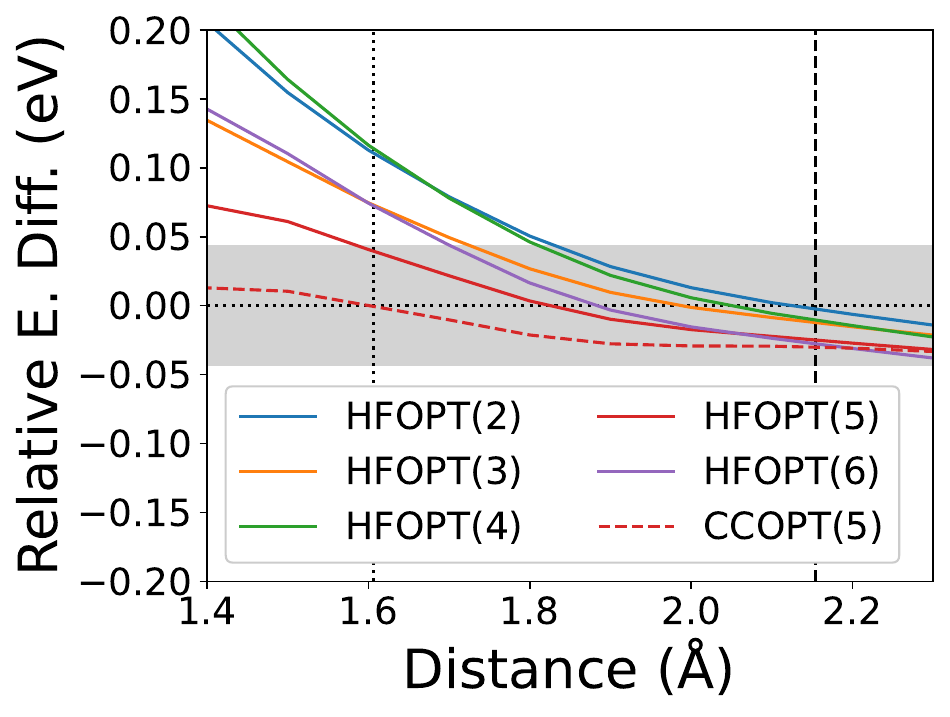} &
    \includegraphics[width=\scale\hsize]{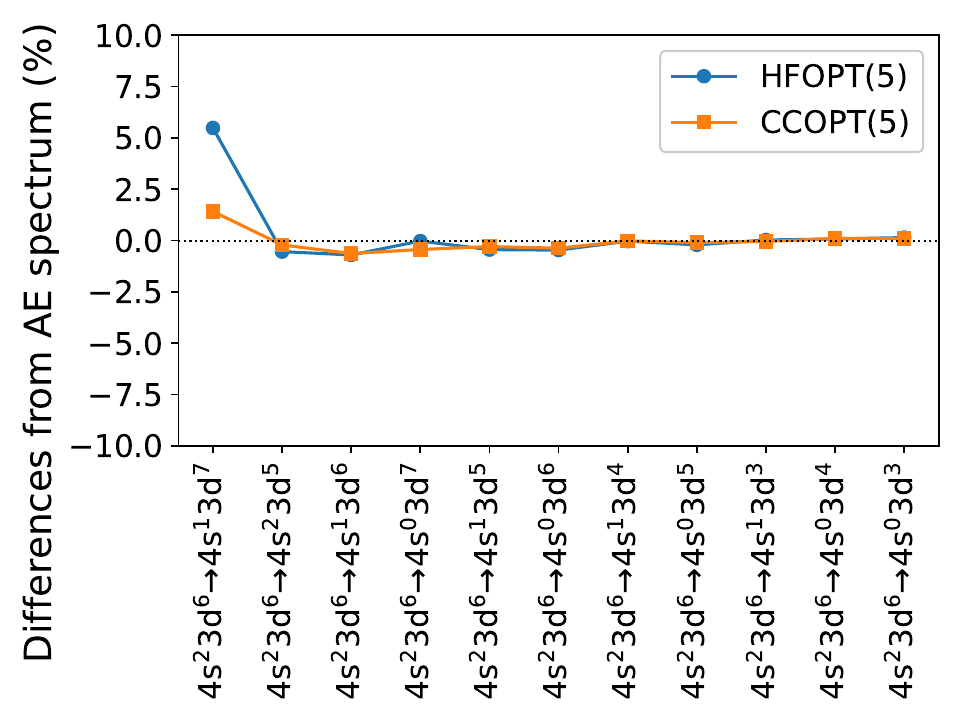} \\
    \raisebox{\offset}{\LARGE Co} &
    \includegraphics[width=\scale\hsize]{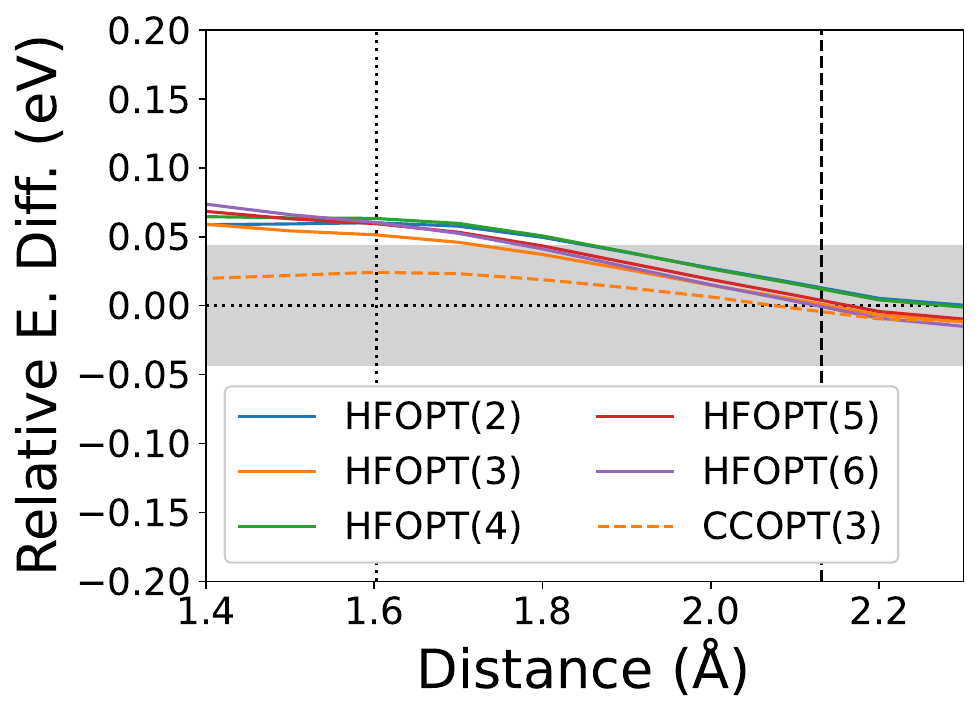} &
    \includegraphics[width=\scale\hsize]{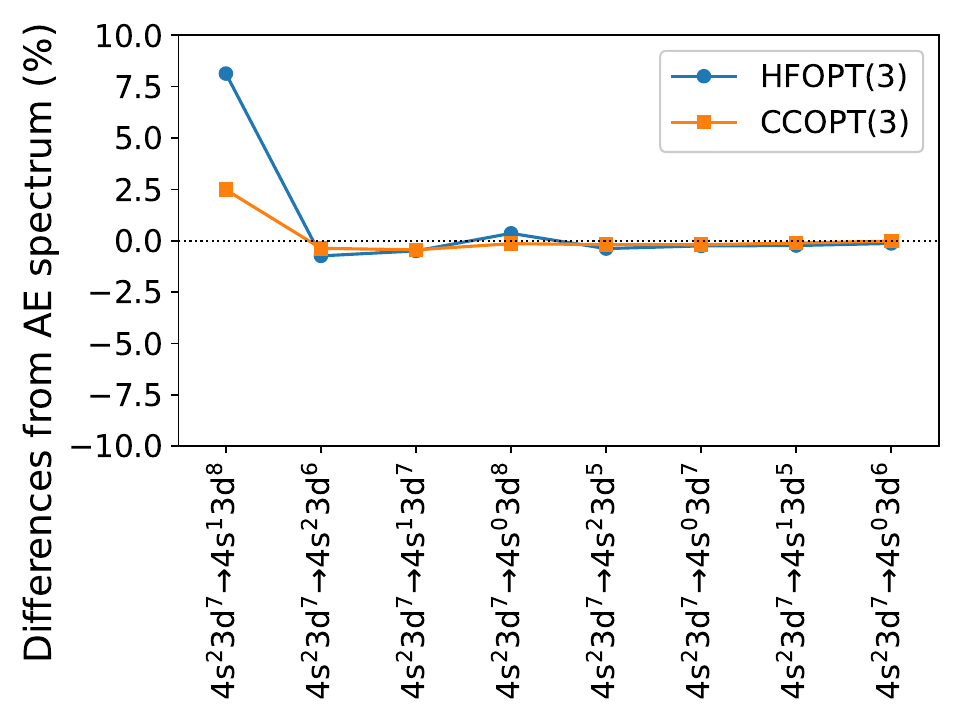} \\
  \end{tabular}
  \caption{
    \textlabel{fig.screening.1}
	(left) 
    Absolute errors of TM--O binding curves (TM=Cr--Co) calculated 
    via CCSD(T) for HFOPTs and CCOPT relative to the AE results. 
    The vertical dashed and dotted lines indicate the AE binding distance 
    and TM–O closest distance in the reference transition metal 
    oxide listed in Table \ref{tab.distance}.    
	(right) 
    Relative errors of the \tadd{energy spectrum} calculated 
    with the best HFOPT and CCOPT relative to the AE results. 
  }
\end{figure*}
\begin{figure*}[htbp]
  \centering  
  \begin{tabular}{rcc}
	& {\Large TM-O binding} & {\Large Ionization} \\
    \raisebox{\offset}{\LARGE Ni} &
    \includegraphics[width=\scale\hsize]{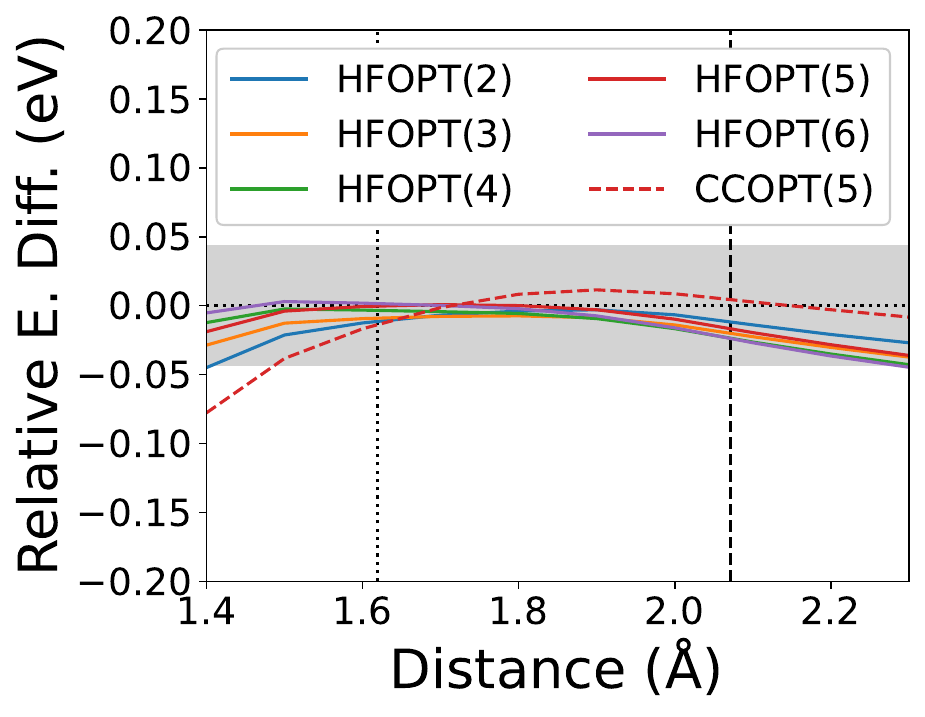} &
    \includegraphics[width=\scale\hsize]{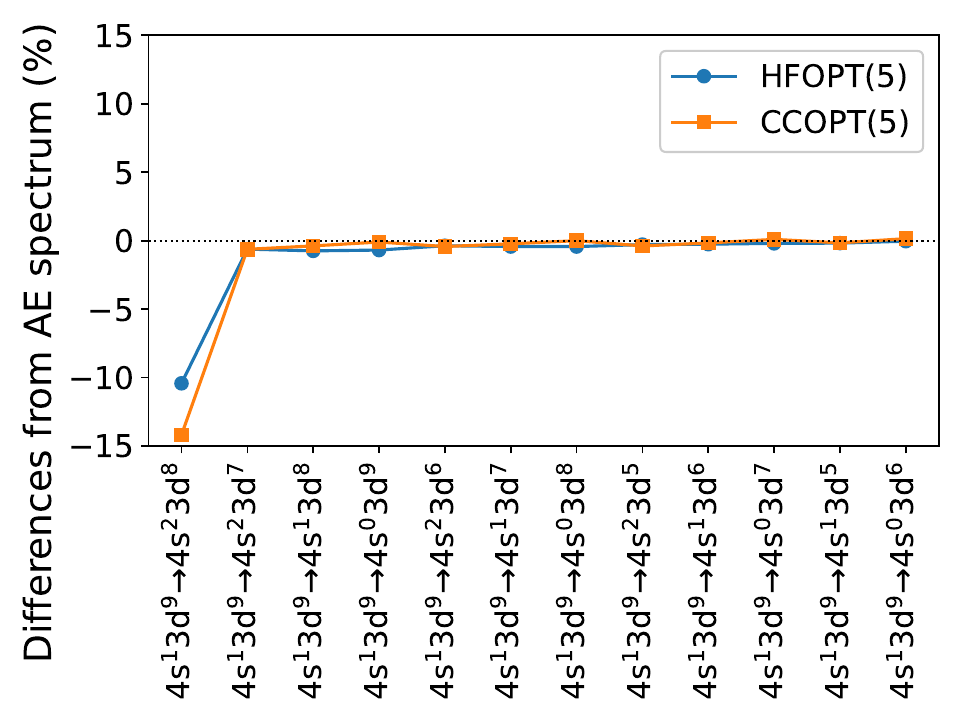} \\
    \raisebox{\offset}{\LARGE Cu} &
    \includegraphics[width=\scale\hsize]{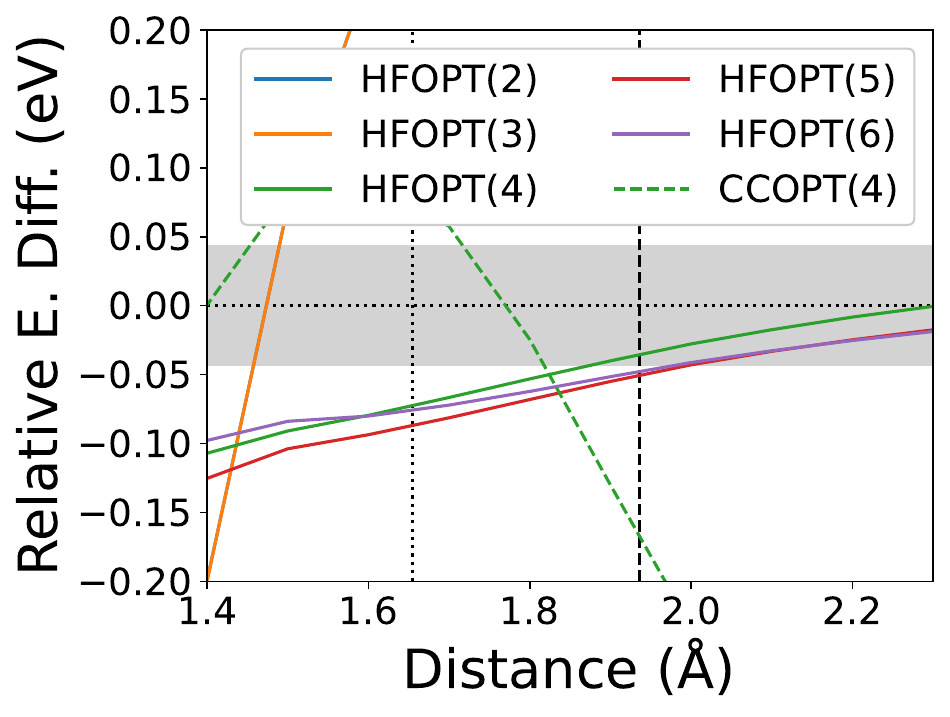} &
    \includegraphics[width=\scale\hsize]{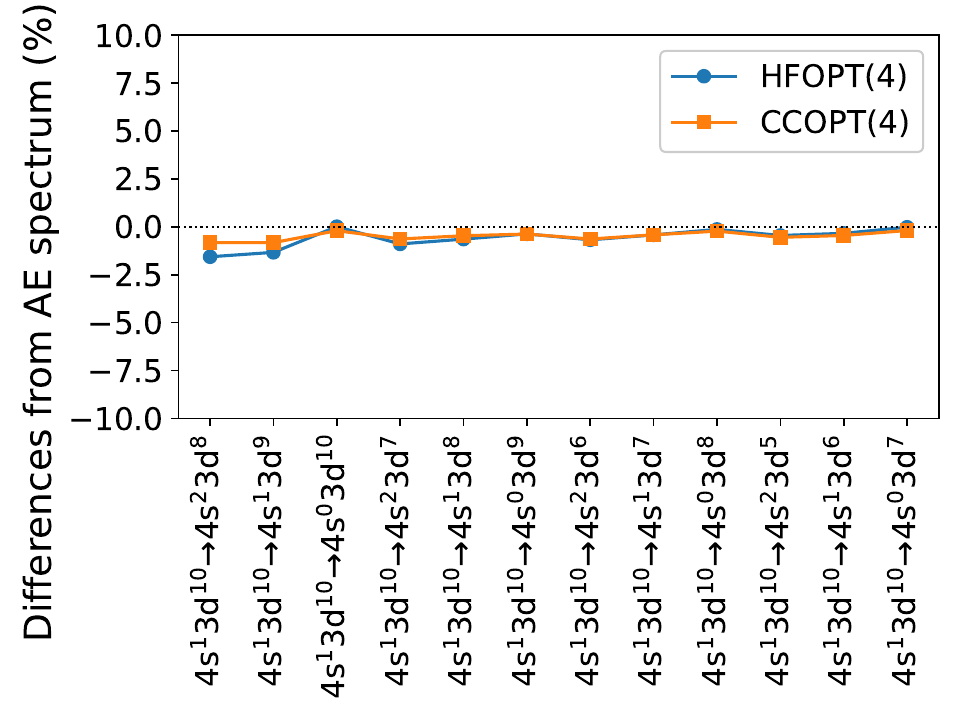} \\
    \raisebox{\offset}{\LARGE Zn} &
    \includegraphics[width=\scale\hsize]{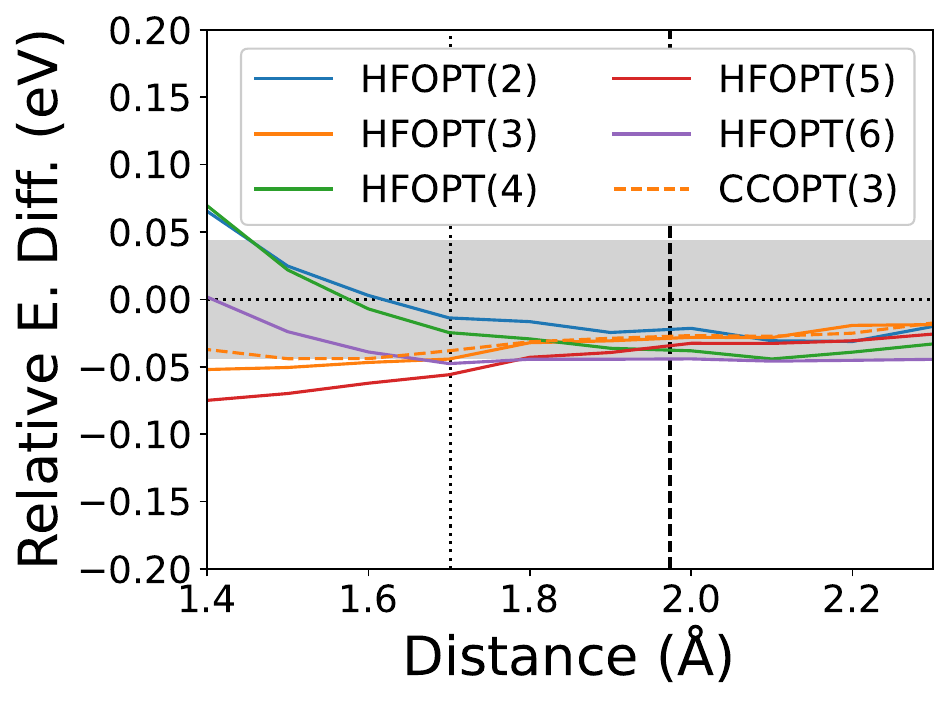} &
    \includegraphics[width=\scale\hsize]{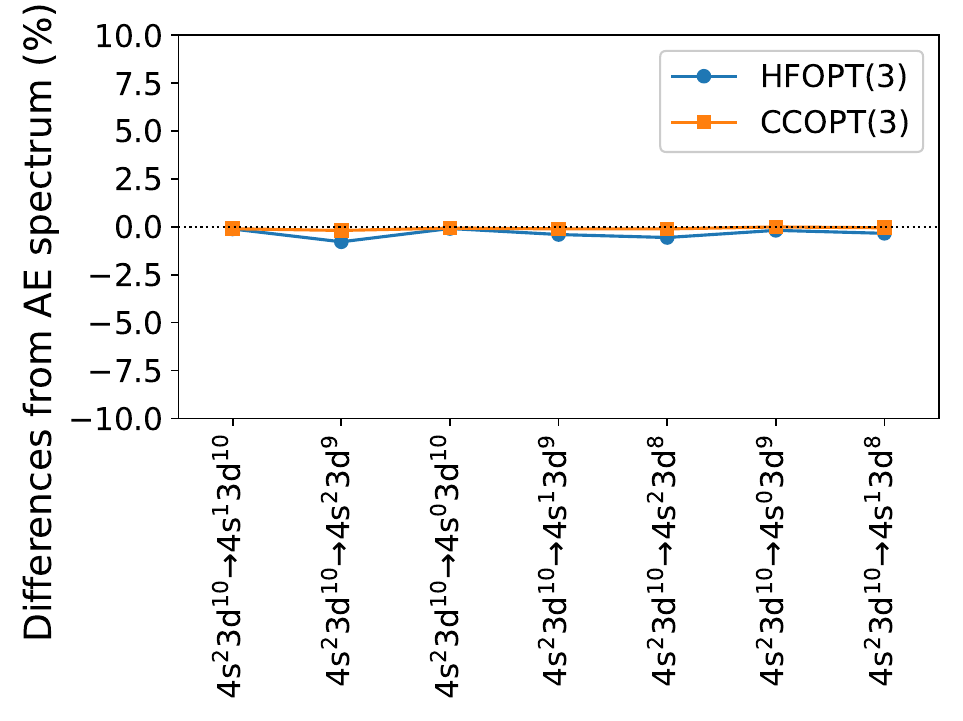} \\
  \end{tabular}
  \caption{
    \textlabel{fig.screening.2}
	(left) 
    Absolute errors of TM--O binding curves (TM=Ni--Zn) calculated 
    via CCSD(T) for HFOPTs and CCOPT relative to the AE results. 
    The vertical dashed and dotted lines indicate the AE binding distance 
    and TM–O closest distance in the reference transition metal 
    oxide listed in Table \ref{tab.distance}.    
	(right) 
    Relative errors of the \tadd{energy spectrum} calculated 
    with the best HFOPT and CCOPT relative to the AE results. 
  }
\end{figure*}
\subsection{
  Validations and comparisons with 
  state-of-the-art semilocal pseudopotentials
}
\textlabel{ssec.comparison}
In the previous Subsection \ref{ssec.screening}, 
we chose a set of PHs {\phname} to be the ones
that best reproduce the AE TM-O binding curves 
and AE ionization energies for each element.
In this Subsection, to test the transferability of {\phname}, 
we investigate their ability to reproduce the AE TM--F binding curve.

\vspace{2mm}
The TM--O and TM--F binding curves obtained 
by our {\phname} and semilocal pseudopotentials,
ccECP \cite{2018AA_LM,2017MCB_LM}, SBKJC \cite{1984WJS_MK,1992WJS_PGJ}, and
BFD \cite{2007MB_MD,2008MB_MD} relative to the AE results are shown in Figure
\ref{fig.verification.1} for Cr--Co and Figure \ref{fig.verification.2} for Ni--Zn. 
The vertical dashed and dotted lines indicate the AE binding distance 
and TM–O/F closest distance in the reference oxide and fluoride listed in Table \ref{tab.distance}. 


\vspace{2mm}
For Mn--F case, 
we used ccECP binding curve as the reference 
instead of AE's due to a convergence issue with the AE calculations. 
For these cases, the binding distance is also taken from the ccECP binding curve. 
Figure \ref{fig.verification.1} shows that ccECP reproduces the AE binding curves 
very well for the other elements, so ccECP can be reasonably used as the reference. 

\vspace{2mm}
\textit{Chromium:}
{\phname} reproduces the AE Cr--O binding curve within chemical accuracy
whereas SBKJC and BFD do not satisfy the chemical accuracy. 
For Cr--F binding curve, {\phname} gives $\sim$1.5 larger deviation than 
the chemical accuracy at the binding distance. 
The deviation gets smaller for longer distances. 
The deviations at the shorter distances are smaller than 
the typical magnitude of locality errors, 0.2--0.3 eV
\cite{2015JY_EE,2015JAS_FAR}. 

\vspace{2mm}
\textit{Manganese:}
{\phname} reproduces the AE Mn--O binding curve almost within 
the chemical accuracy and reproduces the ccECP Mn--F binding 
curve within the chemical accuracy for distances longer than 1.5 \AA. 

\vspace{2mm}
\textit{Iron:}
{\phname} reproduces the AE Fe--O and Fe--F binding curves 
within the chemical accuracy. On the other hand, 
SBKJC (BFD) does not satisfy the chemical accuracy 
for Fe--O (Fe--F) binding curve, respectively. 

\vspace{2mm}
\textit{Cobalt:}
{\phname} reproduces the AE Co--O and Co--F binding curves 
within the chemical accuracy. BFD gives significantly larger 
deviations for both Co--O and Co--F binding curves. 
This is also found for Ni, Cu, and Zn. 

\vspace{2mm}
\textit{Nickel:}
{\phname} reproduces the AE Ni--O binding curve within the chemical accuracy. 
{\phname} gives slightly larger deviations for the Ni--F binding curve 
but they are significantly smaller than the typical magnitude of locality errors, 
0.2--0.3 eV \cite{2015JY_EE,2015JAS_FAR}. 

\vspace{2mm}
\textit{Copper:}
{\phname} gives at most 0.1 eV deviation for the binding curves.
Nevertheless, the deviations are significantly 
smaller than the typical magnitude of locality errors, 
0.2--0.3 eV\cite{2015JY_EE,2015JAS_FAR}. 

\vspace{2mm}
\textit{Zinc:}
{\phname} reproduces the AE Zn--O and Zn--F binding curves 
almost within the chemical accuracy. 

\vspace{2mm}
In summary, {\phname}s reproduce the TM--O and TM--F binding 
curves within or close to chemical accuracy. 
Even when the deviations do exceed chemical accuracy, 
they are significantly smaller than the typical magnitude 
of locality errors, 0.2--0.3 eV\cite{2015JY_EE,2015JAS_FAR}. 
TM-F binding curve is never used to select the {\phname}s. 
Therefore, the adequate reproducibility of the AE TM-F binding 
curve provides assurances on the transferability of the {\phname}s.

\begin{figure*}[htbp]
  \centering  
  \begin{tabular}{rcc}
	& {\Large TM--O binding} & {\Large TM--F binding} \\
    \raisebox{\offset}{\LARGE Cr} &
    \includegraphics[width=\scale\hsize]{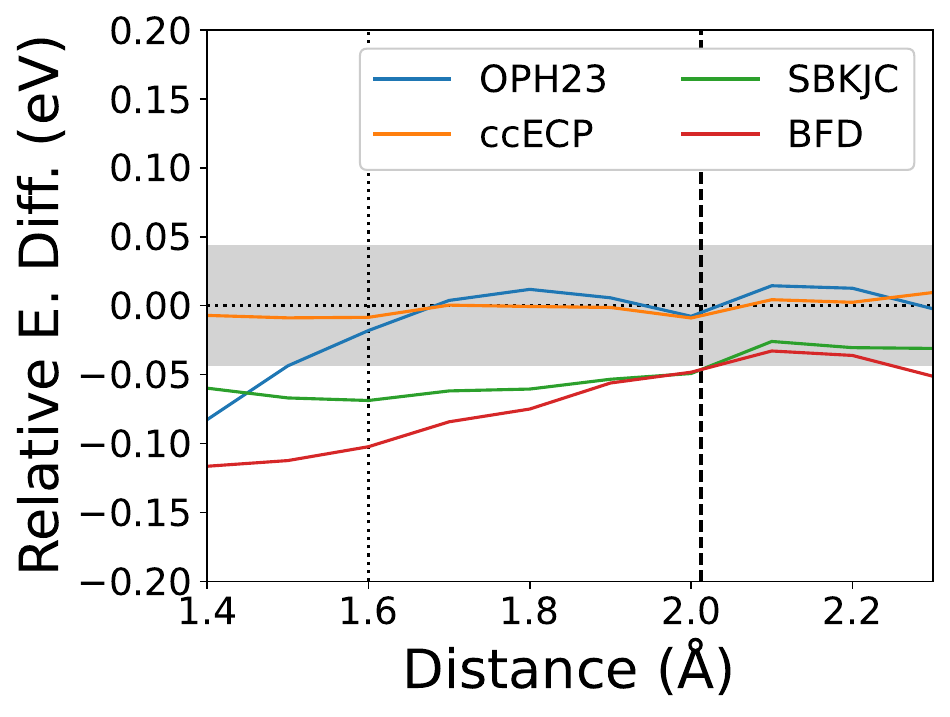} &	
	\includegraphics[width=\scale\hsize]{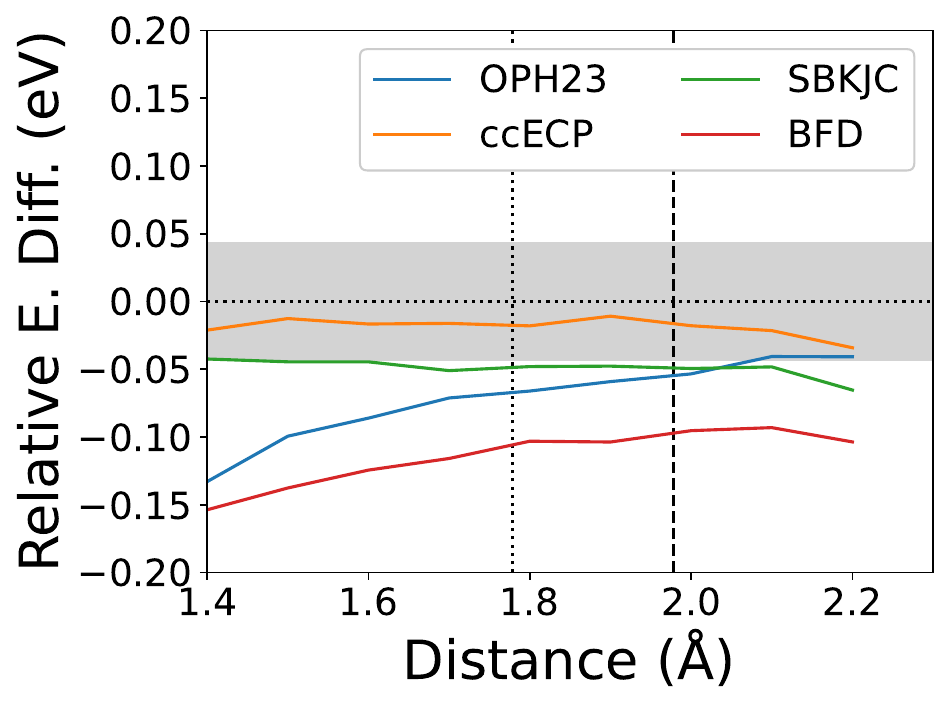} \\
    \raisebox{\offset}{\LARGE Mn} &
    \includegraphics[width=\scale\hsize]{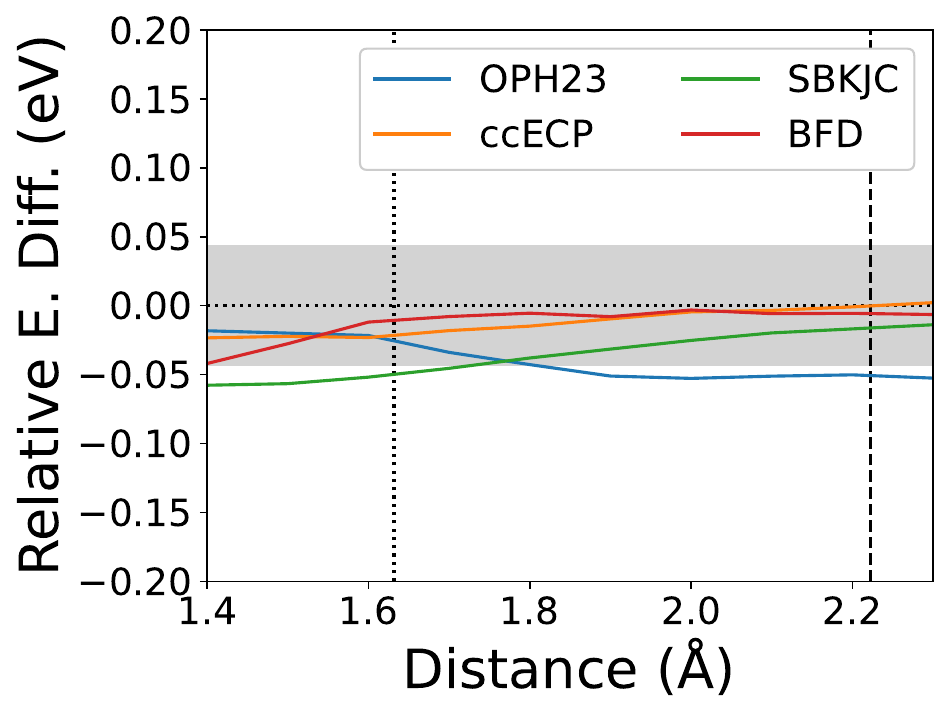} &	
	\includegraphics[width=\scale\hsize]{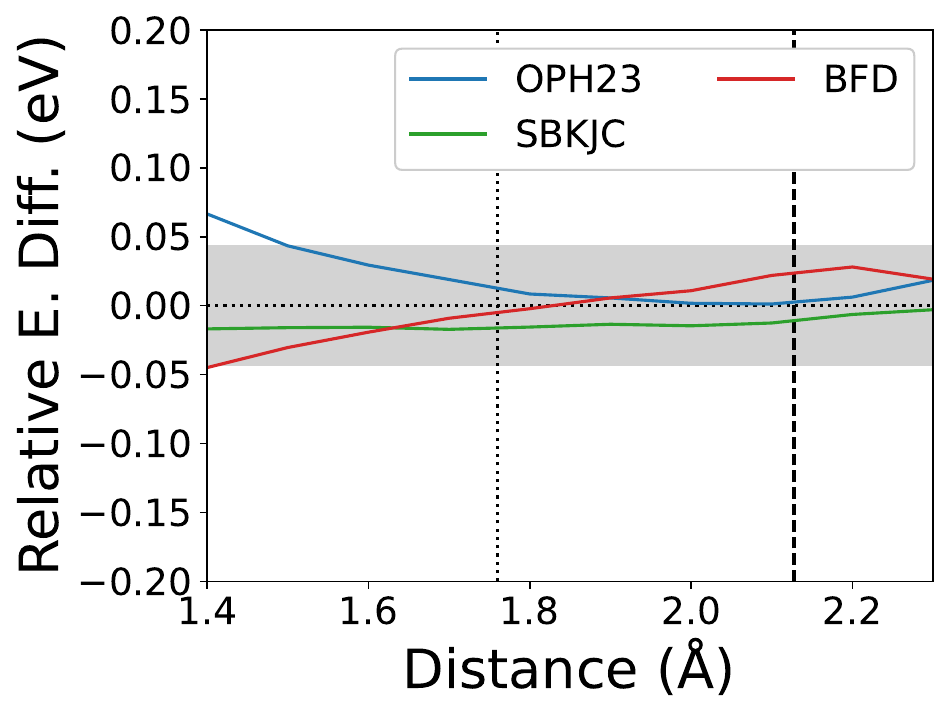} \\
    \raisebox{\offset}{\LARGE Fe} &
    \includegraphics[width=\scale\hsize]{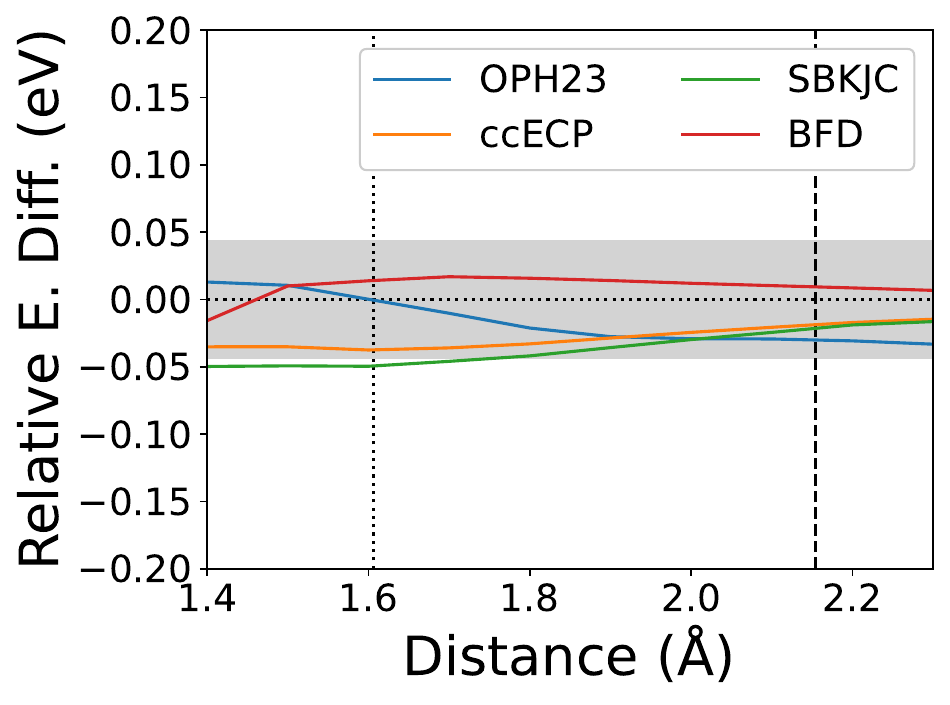} &	
	\includegraphics[width=\scale\hsize]{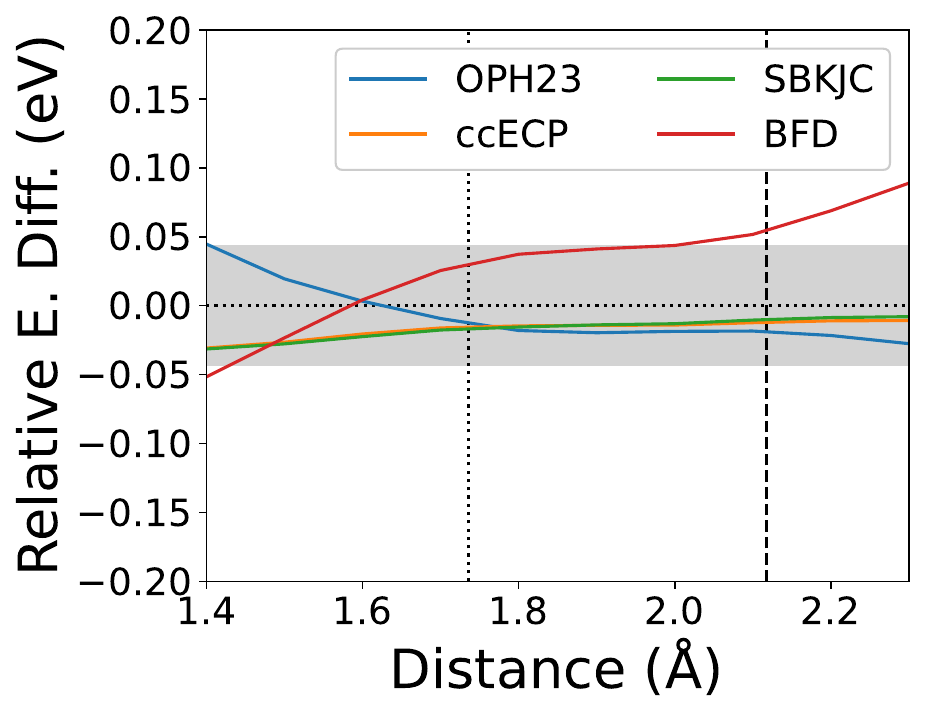} \\
    \raisebox{\offset}{\LARGE Co} &
    \includegraphics[width=\scale\hsize]{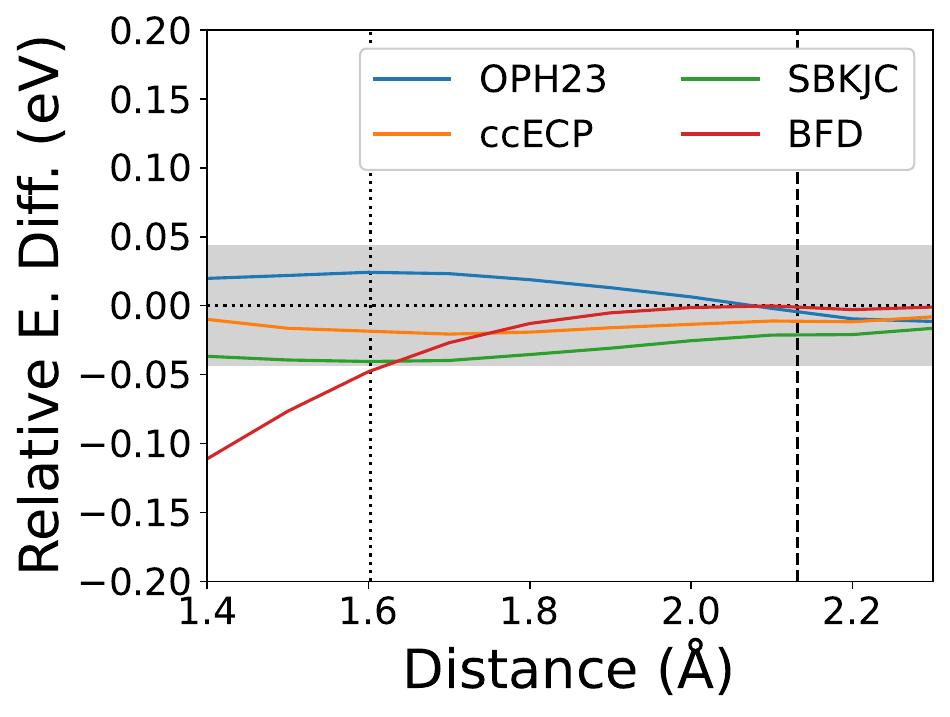} &	
	\includegraphics[width=\scale\hsize]{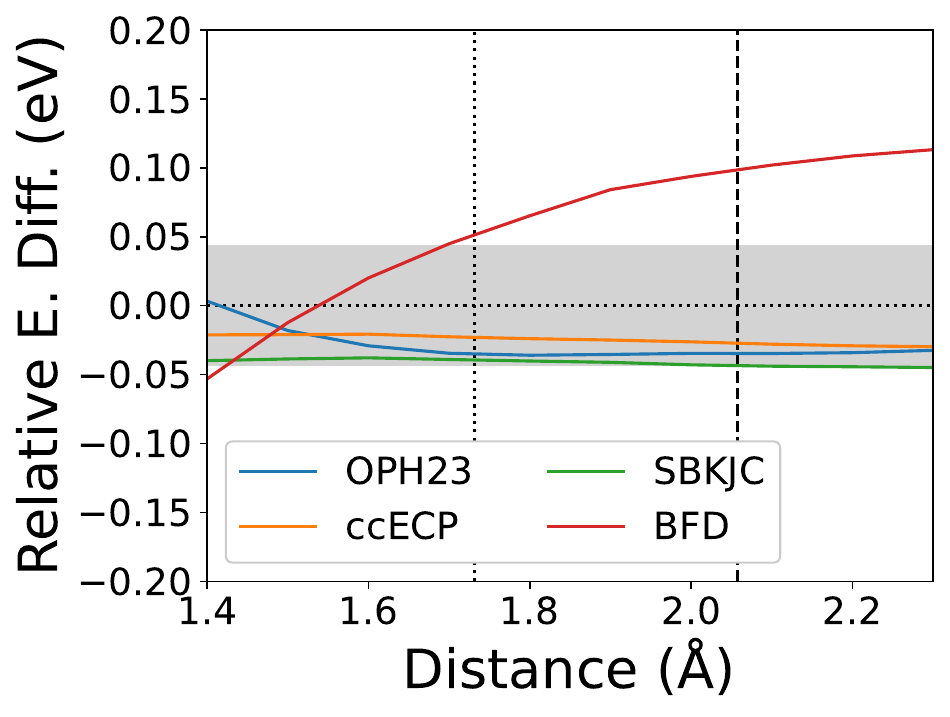} \\
  \end{tabular}
  \caption{
    \textlabel{fig.verification.1}
    Binding curves of TM--O and TM--F molecules (TM=Cr--Co) for {\phname}
    and select community ECPs relative to the AE CCSD(T) results. 
    For the Mn--F case, the binding curves are drawn relative to the ccECP results, 
    because the AE calculations do not converge. 
    The vertical dashed and dotted lines indicate the AE binding distance
    (ccECP is used for the Mn--F case)
    and TM--O(F) closest distance in the reference transition metal 
    oxide (fluoride) listed in Table \ref{tab.distance}.
  }
\end{figure*}
\begin{figure*}[htbp]
  \centering  
  \begin{tabular}{rcc}
	& {\Large TM--O binding} & {\Large TM--F binding} \\
    \raisebox{\offset}{\LARGE Ni} &
    \includegraphics[width=\scale\hsize]{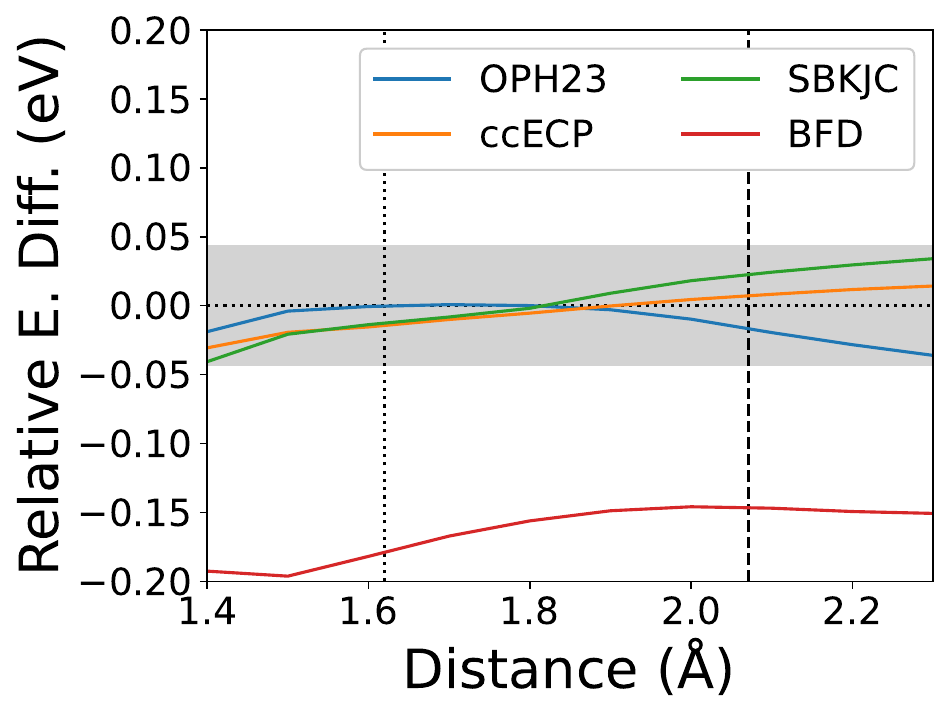} &	
	\includegraphics[width=\scale\hsize]{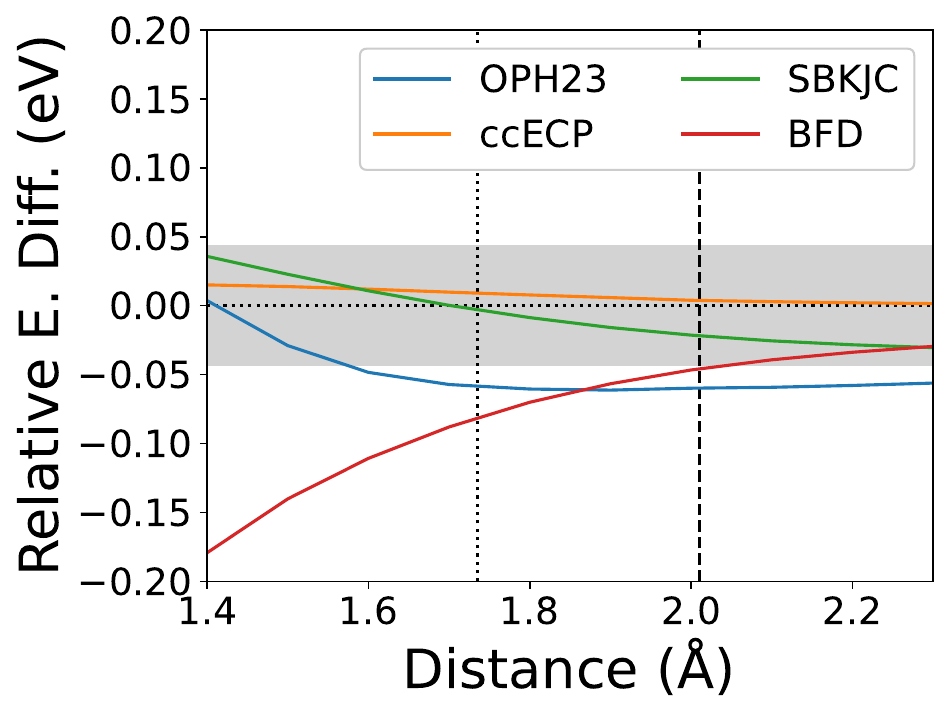} \\
    \raisebox{\offset}{\LARGE Cu} &
    \includegraphics[width=\scale\hsize]{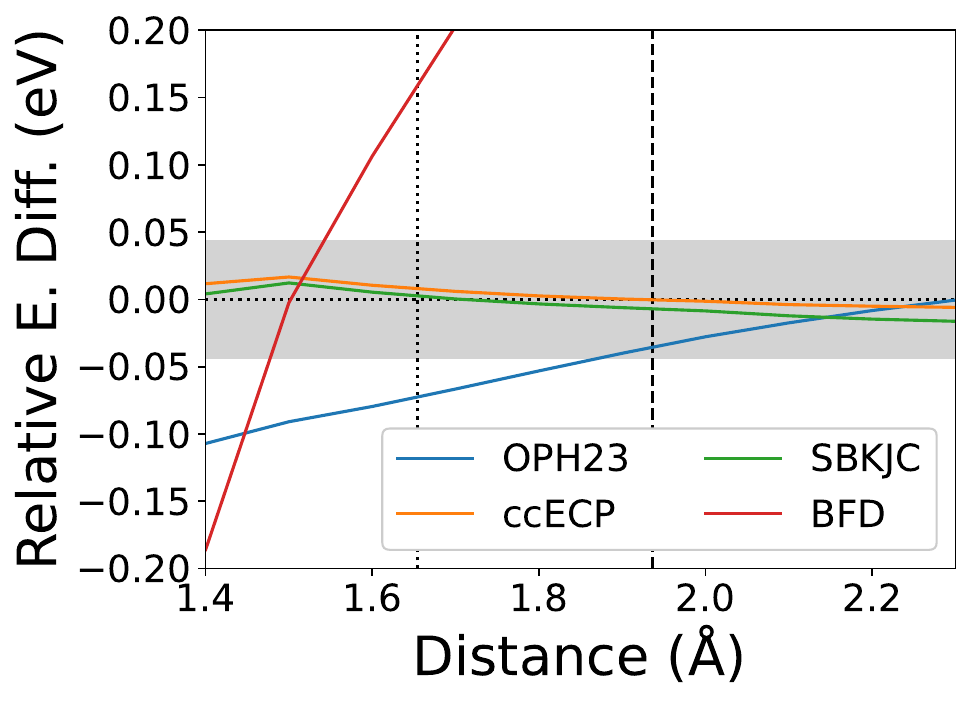} &	
	\includegraphics[width=\scale\hsize]{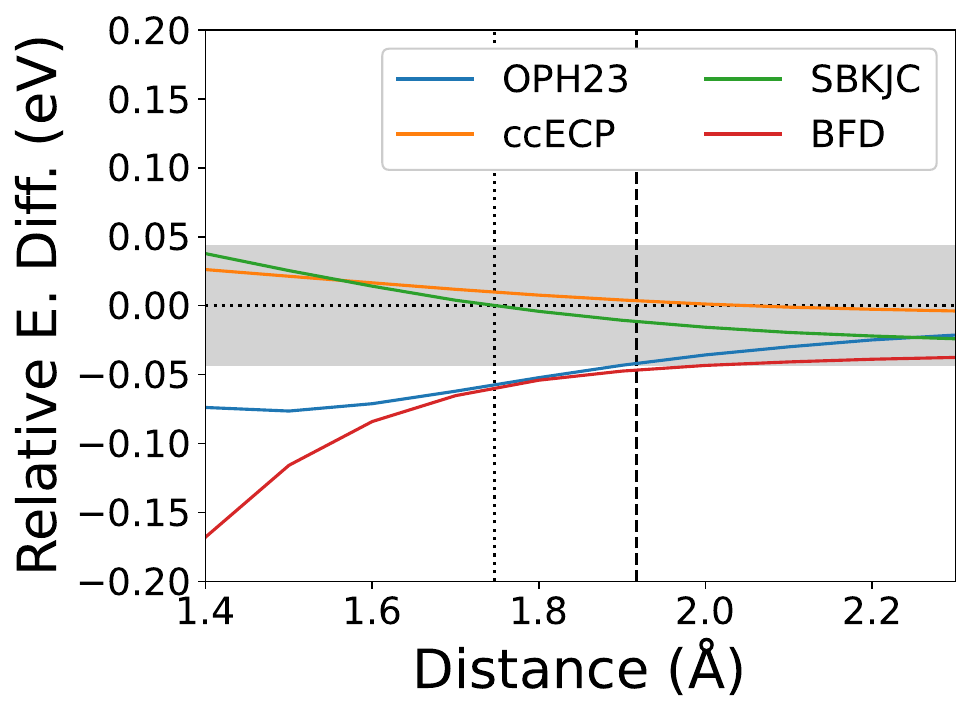} \\
    \raisebox{\offset}{\LARGE Zn} &
    \includegraphics[width=\scale\hsize]{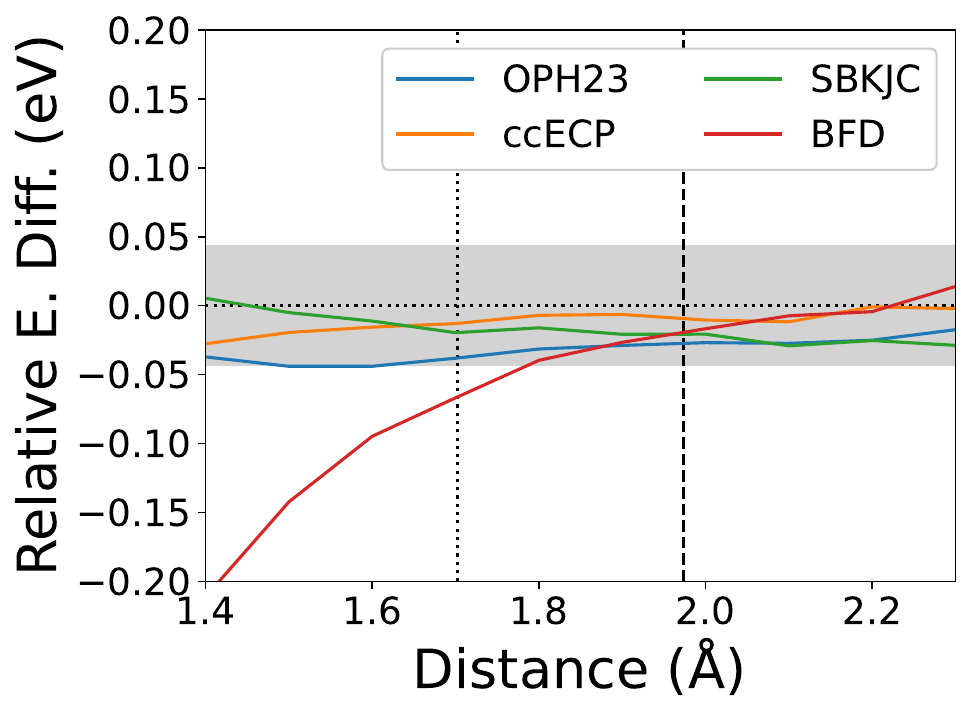} &	
	\includegraphics[width=\scale\hsize]{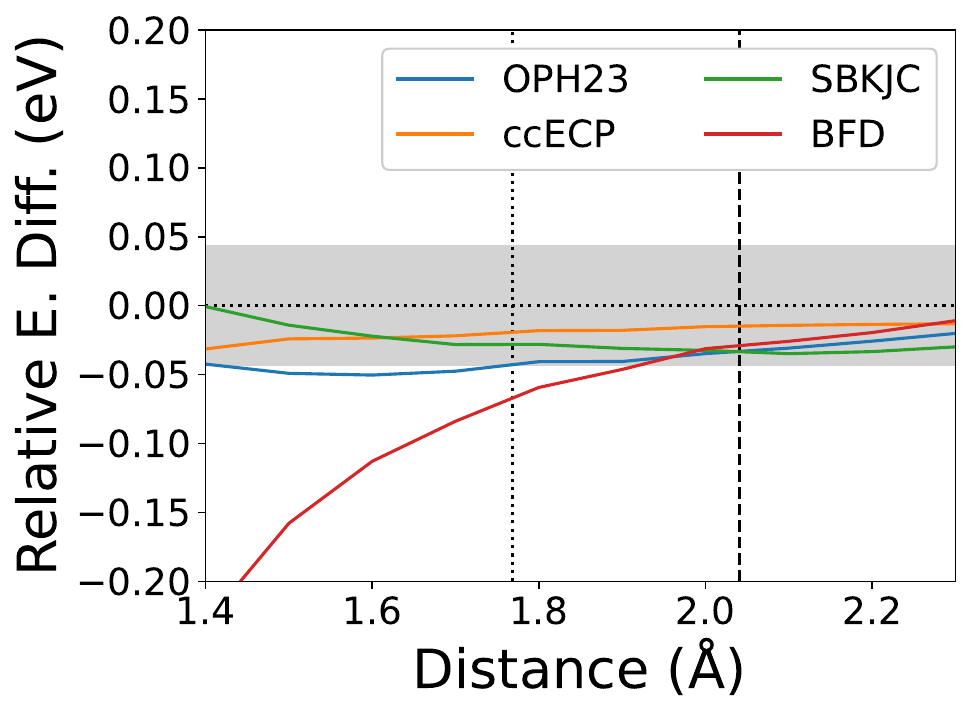} \\
  \end{tabular}
  \caption{
    \textlabel{fig.verification.2}
    Binding curves of TM--O and TM--F molecules (TM=Ni--Zn) 
    for {\phname} and select community ECPs relative to the AE CCSD(T) results. 
    The vertical dashed and dotted lines indicate the AE binding distance 
    given by AE calculations and TM--O(F) closest distance 
    in the reference transition metal oxide (fluoride) 
    listed in Table \ref{tab.distance}.
  }
\end{figure*}
\section{conclusion}\textlabel{sec.conclusion}
We have developed a set of \tadd{PH}s ({\phname}) 
for the Cr--Zn transition metal elements. 
We have used a revised procedure from our previous work\cite{2022MCB_JTK}. 
This procedure can be used for other elements in the periodic table.
\tadd{
  The set {\phname}s are trained to reproduce 
  all the valence orbital norms and eigenenergies 
  with the HF method as well as the neutral excitation 
  and ionization energy spectrum in 3$d$ and 4$s$ levels
with the HF or CCSD(T) method
  using accurate semilocal pseudopotentials, 
  and the AE TM--O binding curve with the CCSD(T) method. 
}
The set {\phname}s reproduced the \tadd{AE} binding curves 
of TM--F within or quite near chemical accuracy. 
\tadd{ 
  The largest deviation from the AE binding curve  
  at the equilibrium distance was 0.067 eV of Cr--F. 
}
The accuracy of the \tadd{PH}s is comparable to 
the best FNDMC pseudopotentials in the literature
\cite{1984WJS_MK,1992WJS_PGJ,2018AA_LM,2017MCB_LM}.
The pseudo-Hamiltionian approach removes the need of 
locality approximations in FNDMC, thus removing locality errors.
Therefore, even when in some cases the deviations 
exceed chemical accuracy, 
they are significantly smaller than the typical magnitude 
of locality errors for transition metal elements, 
0.2--0.3 eV \cite{2015JY_EE,2015JAS_FAR}.

\vspace{2mm}
Our aim is, in the future, to employ the use of {\phname}s 
to delve into an in-depth examination of strongly correlated 
transition metal complexes, focusing on their magnetic structures 
and other intrinsic properties.
While the \tadd{PH}s were designed to reproduce 
the AE ionization energies and AE TM-O binding curves, 
they are found to reproduce the AE TM-F binding curves. 
Further test in solids are required to assess their transferability. 
This will be and addressed in future works. 

\section*{Supporting Information}
\tadd{
	We include the following data as supplementary materials:
    (1) The {\phname} set in XML format compatible with QMCPACK,
    (2) Truncated form of the {\phname} set in the compatible formats with Molpro, Gaussian, and GAMESS, 
    (3) Input and output files of calculations for the binding curves and ionization energies, and
    (4) A PDF file discussing the basis set errors relative to the CBS limit.
}

\section*{Data Availability Statement}
\tadd{
  The data presented in this study are openly available as the supplementary materials. 
}

\section*{Acknowledgments}
The work in ORNL (TI, CB, JK and FR) was supported 
by the U.S. Department of Energy, Office of Science, 
Basic Energy Sciences, 
Materials Sciences and Engineering Division.  
K.H. is grateful for financial support from MEXT-KAKENHI, 
Japan (JP19K05029, JP21K03400, JP22H02170, and JP23H04623), 
and the Air Force Office of Scientific Research, 
United States (Award Numbers: FA2386-22-1-4065).
R.M. is grateful for financial supports from 
MEXT-KAKENHI (JP22H05146, JP21K03400 and JP19H04692), 
from the Air Force Office of Scientific Research 
(AFOSR-AOARD/FA2386-17-1-4049;FA2386-19-1-4015), 
and from JSPS Bilateral Joint Projects (JPJSBP120197714). 
An award of computer time was provided by the Innovative 
and Novel Computational Impact
on Theory and Experiment (INCITE) program. 
This research used computational resources of the Oak Ridge 
Leadership Computing Facility, 
which is a DOE Office of Science User Facility supported 
under Contract DE-AC05-00OR22725. 
This research also used resources of the Research Center
for Advanced Computing Infrastructure (RCACI) at JAIST.

\section*{Appendix}\textlabel{sec.appendix}
\tadd{
\subsection{Angular momentum eigenfunctions}
\label{appendix.sh}
In the position representation with 
the spherical coordinates $(r, \theta, \phi)$, 
the eigenfunctions for the angular momentum 
eigenstates $\left|\ell m\right\rangle$ 
are given by the scalar spherical harmonics:
\begin{equation}
Y_{\ell}^{m}\left(\theta,\phi\right)=\left(-1\right)^{\left(m+\left|m\right|\right)/2}\sqrt{\frac{2\ell+1}{4\pi}\frac{\left(\ell-\left|m\right|\right)!}{\left(\ell+\left|m\right|\right)!}}P_{\ell}^{\left|m\right|}\left(\cos\theta\right)^{im\phi}.
\end{equation}
Here, $P_{\ell}^{\left|m\right|}(t)$ is 
the associated Legendre polynomials:
\begin{equation}
P_{\ell}^{m}\left(t\right)=\frac{1}{2^{\ell}}\left(1-t^2\right)^{\frac{m}{2}}\sum_{j=0}^{\left[\left(\ell-m\right)/2\right]}\frac{\left(-1\right)^{j}\left(2\ell-2j\right)!}{j!\left(\ell-j\right)!\left(\ell-2j-m\right)!}t^{\ell-2j-m}.
\end{equation}
Here, $\left[\left(\ell-m\right)/2\right]$ indicates 
the maximum integer less than or equal to $(\ell-m)/2$.
}

\subsection{Justification of the many-body correction}
\label{appendix.mbc}
The difference of excitation energies between HF and CCSD(T) 
almost does not depend on the choice of ECP \cite{2022MCB_JTK,2018AA_LM} so 
\begin{equation}
  \left(\Delta E_i^{\mathrm{CCSD(T),CCOPT}} - \Delta E_i^{\mathrm{HF,CCOPT}}\right) \approx \left(\Delta E_i^{\mathrm{CCSD(T),HFOPT}} - \Delta E_i^{\mathrm{HF,HFOPT}}\right).
  \label{eq.indepedence}
\end{equation}
Thus, the numerator in the parentheses of \eqref{eq.ccopt} is approximated as 
\begin{equation}
  \Delta E_i^{\mathrm{CCSD(T),CCOPT}} - \Delta E_i^{\mathrm{CCSD(T),ccECP}}.
\end{equation}
Therefore, the minimization of the cost function in Eq. \eqref{eq.ccopt} 
corresponds to the optimization so as to reproduce the electron excitation energies 
at the CCSD(T) level. 


\bibliography{references}
\end{document}